\documentclass{llncs}
\usepackage{graphicx}
\usepackage{subfigure,tikz,bm}
\usepackage{amsmath}
\usepackage{amsfonts}
\usepackage{amssymb}
\usepackage{todonotes,tikz}
\usepackage{xcolor}
\usepackage{graphicx}
\usepackage{hyperref}
\usepackage[outdir=./]{epstopdf}
\usepackage{algorithm}
\usepackage{algorithmic}
\usepackage{float}  
\usepackage{lipsum}
\usepackage{ wasysym }
\usepackage{booktabs}
\usepackage{array}

\usepackage{mathtools,nccmath}
\usepackage{times}

\usepackage{multirow}
\usepackage{wrapfig}
\usetikzlibrary{shapes,snakes}
\makeatletter
\newenvironment{breakablealgorithm}
{
	\begin{center}
		\refstepcounter{algorithm}
		\hrule height.8pt depth0pt \kern2pt
		\renewcommand{\caption}[2][\relax]{
			{\raggedright\textbf{\ALG@name~\thealgorithm} ##2\par}%
			\ifx\relax##1\relax 
			\addcontentsline{loa}{algorithm}{\protect\numberline{\thealgorithm}##2}%
			\else 
			\addcontentsline{loa}{algorithm}{\protect\numberline{\thealgorithm}##1}%
			\fi
			\kern2pt\hrule\kern2pt
		}
	}{
		\kern2pt\hrule\relax
	\end{center}
}
\makeatother

\usepackage{cleveref}

\RequirePackage{todonotes}

\long\def\comment#1{}

\newcommand{\oomit}[1]{}
\newcommand{\define}{\widehat{=}}

\AtBeginDocument{%
  \providecommand\BibTeX{{%
    \normalfont B\kern-0.5em{\scshape i\kern-0.25em b}\kern-0.8em\TeX}}}

\begin{document}

\title{Correct-by-Construction for Hybrid Systems by Synthesizing  Reset Controller}
\author{}
\institute{}
\maketitle

\begin{abstract} 
Controller synthesis, including reset controller, feedback controller, and switching logic controller, provides an essential mechanism to guarantee the correctness and reliability of hybrid systems in a correct-by-construction manner. Unfortunately, reset controller synthesis is still in an infant stage in the literature, although it makes theoretical and practical significance. In this paper, we propose a convex programming based method to synthesize reset controllers for polynomial hybrid systems subject to safety, possibly together with liveness. Such a problem essentially corresponds to computing an initial set of continuous states in each mode and a reset map associated with each discrete jump such that any trajectory starting from any computed initial state keeps safe if only safety constraints are given or reaches the target set eventually and keeps safe before that if both safety and liveness are given,  through the computed reset maps. Both cases can be reduced to reach-avoid and/or differential invariant generation problems, further encoded as convex optimization problems. Finally, several examples are provided to demonstrate the efficiency and effectiveness of our method.
\end{abstract}


 \section{Introduction}
Controller synthesis, including reset controller, feedback controller, and switching logic controller, provides an essential mechanism to guarantee Hybrid systems (HSs), nowadays also known as cyber-physical systems (CPSs), exploit networked computing units to monitor and control physical processes via wired and/or radio communications, essentially combine discrete mode changes with the continuous evolution, routinely described by differential equations. HSs are omnipresent in our daily life, from spacecraft to high-speed train control systems, to power and control grids, to automated plants and factories, to name just a few. Many HSs are entrusted with mission- and/or safety-critical tasks. Therefore, efficient and verified development of safe and reliable HSs is a priority mandated by many standards, yet a notoriously challenging domain.

Controller synthesis, given a model of the assumed behaviour of the environment and a system goal, algorithmically constructs an operational behaviour model for a component that, when executing in an environment consistent with the assumptions, results in a system that is guaranteed to satisfy the goal. Controller synthesis provides a correct-by-construction manner for developing reliable HSs, which has attracted increasing attention from computer science and control theory in the past decades. In HSs (CPSs), operation (i.e., control) could be inputs to  differential equations, or switch conditions from one mode to another one,  or initial conditions for each mode and reset maps when conducting discrete jumps. So,  controllers can be naturally classified into three categories, namely, feedback controllers, switching logic controllers, and reset controllers. In the literature, there are huge bulk of work on the synthesis of the first two types of controllers, please refer to  
\cite{tomlin2000,Asarin00,coogan2012guard,jha2010synthesizing,2009SwitchedSystem,Girard12,GulwaniT08,2009SwitchedSystem,ZhaoZK13} and 
the references therein. However, the synthesis problem of 
the third type is still a virgin land, although reset controller synthesis  is not only theoretically significant but also makes important sense in practice, as many important practical problems can be reduced to reset controller synthesis, e.g.,  the substantial instantaneous change in velocity of a spacecraft induced by impulsive controls in satellite rendezvous \cite{brentari2018hybrid}, also re-configuring safety-critical devices like spacecrafts when an exception happens, and so on. What's more, as indicated by the following motivating example, in some cases, only with feedback and switching logic controllers, even their combination, without reset controller,  one cannot achieve the system goal.

\begin{example}[A Motivating Example] \label{ex:running}
Consider the hybrid system  given in Fig.~\ref{fig:intro1}. 
{\small 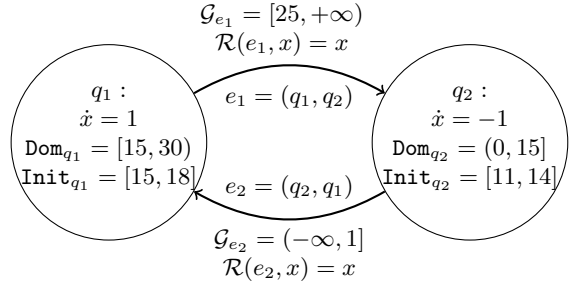
\begin{figure}[h]
\vspace{-0.5cm}
\centering
\scalebox{1}{  \begin{tikzpicture}
  \centering
\tikzstyle{mode} = [circle, minimum size=2.6cm,inner sep=0pt, text centered, draw=black, fill=white!70]
\node (mode1) [mode, align=center] {};
\node (mode2) [mode, right of=mode1, xshift=3.8cm] {};
\draw [->,thick]  (mode1) to  [bend left] node[anchor=south, yshift=0.2cm] {} (mode2);
\draw [->,thick] (mode2) to [bend left] node[anchor=south,yshift=-0.7cm]{} (mode1);
\node [above of=mode1, yshift=-0.9cm] {$ \begin{array}{c}
    q_1 :  \\
  \dot{x} =1  \\
\mathtt{Dom}_{q_1} =  [15, 30) \\
\mathtt{Init}_{q_1} = [15,18]
 \end{array}$};
 \node[above of=mode1,xshift=2.3cm,yshift=0.5cm]{$ \begin{array}{c}
    \mathcal{G}_{e_1}=[25,+\infty)\\
    \mathcal{R}(e_1,x)=x
 \end{array}$};
  \node[above of=mode1,xshift=2.4cm,yshift=-0.4cm]{$e_1=(q_1,q_2) $};
 \node[above of=mode1,xshift=2.4cm,yshift=-2.5cm]{$ \begin{array}{c}
    \mathcal{G}_{e_2}=(-\infty,1]\\
    \mathcal{R}(e_2,x)=x
 \end{array}$};
  \node[above of=mode1,xshift=2.4cm,yshift=-1.6cm]{$e_2=(q_2,q_1)$};
 \node [above of=mode2, yshift=-0.9cm] {$ \begin{array}{c}
    q_2 :  \\
  \dot{x} =-1  \\
  \mathtt{Dom}_{q_2}=(0,15] \\
  \mathtt{Init}_{q_2}=[11,14]
 \end{array}$};
\end{tikzpicture}}
\caption{Hybrid Automaton for Example~\ref{ex:running}}
\label{fig:intro1}
\end{figure} }
\oomit{
\begin{figure}[t]
\centering
\subfigure[]{
\includegraphics[width=3.2cm]{figure_new/intro1.eps}
}
\hspace{-3mm}
\subfigure[]{
\includegraphics[width=4cm]{figure_new/intro2.eps}
}
\caption{Dynamic of the Automata
}
\label{fig:intro2}
\end{figure}}
\vspace{-0.3cm}
Suppose the safe sets in $q_1$ and $q_2$ are $\mathcal S_1  = [15,31)$ , $\mathcal S_2  = (0,14]$, respectively. Firstly, as the dynamics in the two modes both are autonomous, it is impossible to have feedback controllers for them to maintain the safety. Secondly, it is not hard to show that no switch control strategy can be applied to guarantee the safety of the system by strengthening the guard conditions and domain constraints either, as once a jump from $q_1$ to $q_2$ or vice versa, the system will arrive in the unsafe set. However, if it is allowed to redesign the reset map $\mathcal{R}$, clearly, one can easily obtain a refined hybrid automaton that satisfies the safety property.  
\end{example}

In this work, we investigate the reset controller synthesis problem for HSs. A system goal is usually a safety constraint, possibly together with a liveness constraint. A reset map associated with a discrete jump between two modes is, in general, a set-valued function that specifies how continuous evolutions in the post-mode are related to the ones in the pre-mode. Reset controller synthesis is to strengthen the initial condition in each mode and to synthesize a reset map for each discrete jump for a considered HS w.r.t. the given system goal, so that the refined HS satisfies the system goal. If only safety constraint is concerned, such a synthesis problem essentially corresponds to compute 
an initial set  associated with each mode and to 
synthesize a reset map associated with each jump such that  
each any continuous flow from 
any computed initial set either reaches to the guard of a jump  eventually, or stay inside the mode subject to the safety constraint forever. If the former 
happens, the synthesized reset map associated with the jump can guarantee the safety in the post-mode of the jump. This problem can be further reduced to a reach-avoid problem. Traditionally, reach-avoid problem solves how to compute the maximal set of initial states such that the considered system starting from which is guaranteed to reach the  target set eventually while remaining inside the safe set until hitting the target. Moreover, we show that the maximal reach-avoid set can be inner-approximated by reduction to a convex programming problem, which could be solved with on-the-shell  SDP solvers. The latter can be reduced to the differential invariant generation problem,  which can be well solved by exploiting existing methods for computing differential invariants, e.g., \cite{LiuZZ11,Ghorbal2014,xue2019hscc,Wang2021}. If safety and liveness properties are considered together, we have to address the following  two problems: 
how to guarantee to reach to the target set in or the must-jump part  of the guard of a jump outgoing from a mode while keeping safe until  reaching the target or leaving the mode via the jump, and how to avoid the unreachability caused by infinite loops among the modes. The must-jump part of the guard of a jump 
means the intersection of the guard and 
the complementation of the domain of the pre-mode of the jump, to which the jump must take place immediately in case a trajectory reaches. The former problem essentially corresponds to a reach-avoid problem, while the latter problem  can be solved by searching all simple loops among the modes and blocking them.

We implement a prototypical tool and provide several examples to demonstrate the effectiveness and performance of the proposed method. 

In summary, the main contributions of this paper include 
\begin{enumerate}
\item We reduce synthesizing safe (with live) reset controllers to 
 reach-avoid and differential invariant generation problems. 
\item We propose to inner-approximate reach-avoid sets by reduction to convex programming problems, which can be efficiently solved  using on-the-shell SDP solvers. 
\item A prototypical implementation is provided and  applied to several case studies
 to illustrate the effectiveness and efficiency of our approach. 
\end{enumerate}

\paragraph{Reset controller synthesis and time-delay} 
Physically, changing a continuous evolution abruptly is counter-intuitive, even impossible, for example, it is impossible to change the velocity of a train from  0 km/h to 300 km /h instantaneously in reality, although it is mathematically simple. Actually, such a reset procedure takes time, even such a procedure can be done through a discrete action, as the actuator itself takes time to make the control into effect. So, ideally, we should consider this issue in the context with time-delay like delay hybrid automata \cite{Bai2021}, so that the time spent by the reset controller can be modeled as time delay and thus it can be taken into account. 
As a heuristic exploration, in this paper, we want to investigate this issue in a simpler mathematical model, and therefore abstract away time-delay caused by reset controllers. However,  we believe that our approach still works in the context of time-delay by exploiting recent results on invariant generation for delay hybrid systems in \cite{Bai2021} and reach-avoid problem for delay differential equations in \cite{Xue2021}, that will be a future work.

\subsection{Related Work}
A natural idea for automatic verification of HSs (CPSs) is state-space
exploration aiming at computing the reachable state space. Unfortunately, the exact computation of reach sets is impossible in general \cite{HKPV98}, especially for systems with nonlinearity, albeit with decidable families of sub-classes (see, e.g.,~\cite{LPY01,Gan18}). A more generally applicable option is to compute over- and under-approximations of the state sets reachable under time-bounded continuous dynamics, and then to embed them, e.g., into depth-bounded automatic verification by bounded model checking, or into unbounded verification by theorem proving. Thus, various abstraction techniques have been proposed for over- and under-approximating reachable sets of continuous dynamics given as ordinary differential equations, e.g., based on interval arithmetic \cite{RatschanS05}, Taylor models \cite{berzmakino:1998,neheretal:2007}, polyhedral \cite{chutinan:1998}, zonotopes \cite{Girard:Zonotopes}, ellipsoids \cite{KurzhanskiVaraiya:Ellipsoids},  and support functions  \cite{LeGuernicGirard:SupportFunctions} and so on,  as well as abstractions based on discovering invariants \cite{LiuZZ11,KongHSHG13,SogokonJJ19,Wang2021} etc. There are several bounded model checkers available for HSs, e.g.,  iSAT-ODE \cite{eggers2008sat}, Flow* \cite{Chenxin13}, and dReach \cite{Gao15}. Theorem provers for HSs are also available, e.g., KeYmaera \cite{Platzer10} or HHL Prover \cite{Zou13,WangZZ17}.

As the two sides of a coin, verification of  HSs can also be conducted in correct-by-construction manner by synthesizing controllers, including synthesizing feedback controllers, switching logic controllers, and reset controllers. In the literature, there is a huge volume of work on synthesizing feedback controllers and switching controllers for HSs, and we just list a few of them below. 

Feedback controllers steer all continuous behaviour away from the unsafe region through computed inputs to change physical disciples of continuous evolution. There are a rich family of methods contributing to this kind of safe controllers synthesis such as moment-based methods (e.g., \cite{zhao2019optimal}), Hamilton-Jacobi based methods (e.g., \cite{tomlin2000}), Lyapunov functions or barrier certificates based methods (e.g., \cite{ames2016control}), abstraction-based methods (e.g., \cite{Paulo0032856,Girard12}), counter-examples guided inductive synthesis methods (e.g., \cite{AbateBCCDKKP17}), etc. Switching logic controllers strengthen the domain constraint for each mode under which continuous evolution is allowed, and the guard is associated with each discrete jump. In the literature, switching logic synthesis  has been extensively studied, and various approaches have been proposed, which can be categorized into abstraction based, e.g., 
\cite{Paulo0032856,BeltaBook,Girard12,Reissig2016,DBLP:journals/deds/NilssonOL17,HSCC2018}, and constraint solving based, e.g., \cite{ZhaoZK13,2009SwitchedSystem,TalyT10}. 

\oomit{The basic idea of abstraction based approaches is to  abstract the original system under consideration to a finite-state two-players  game, and  then solve   reactive  synthesis  using automata-theoretic  algorithms  with  respect  to temporal  control  objectives. In contrast, 
the basic idea of  constrains solving based approaches is to  reduce the synthesis problem to 
an invariant generation problem, which can be further reduced to a constraint solving problem. 
As a generalization of \cite{TalyT10},  an optimal switching controller synthesis 
is investigated  in \cite{jha2010synthesizing} by solving an unconstrained numerical optimization problem.
Based on reachable set computation and fixed point iteration, a general framework of controller synthesis for HSs is proposed in \cite{871306,871303}.}
 
While reset controller defines a set of initial states associated with each mode and a mapping associated with each discrete jump that maps states in the pre-mode satisfying the guard to states in the post mode so that any trajectory of the considered HS meets the system goal. The problem was investigated by Clegg in \cite{clegg1958nonlinear} to overcome the limitations of linear control, which provides the feedback control system with a reset structure. Most of the existing work mainly focuses on stability analysis of linear systems, please refer to \cite{Beker1999,guo2009stability,prieur2018analysis} and the references therein. 

In contrast to existing work, in this paper, we present automated synthesis algorithms by computing inner-approximations of reach-avoid sets and/or differential invariant sets to synthesize an initial set associated with each mode and a reset map associated with each jump for a given HS  s.t. the resulting HS respects the system goal. 


The remainder of this paper is structured as follows. In Sect.~\ref{pre}, we introduce the notions of HSs and the problems of interest, and develop some necessary theories. Sect. \ref{sec:reset_map_synthesis} presents the main framework of reset controller synthesis, and Sect. \ref{sec:SDPavoid} focus on the implementation based on SDP. Sect.~\ref{sec:ier} reports implementation and experiments, and 
we conclude this paper in Sect.\ref{sec:conclusion}. 
 \section{Preliminaries}
\label{pre}

Throughout this paper, we use $\mathbb{R}$, $\mathbb{Z}$ and 
$\mathbb{N}$ to denote the set of real, integer and natural numbers, respectively. 
$\mathbb{R}^n$ is the set of $n$-dimensional real vectors. For a given set $S$, $P(S)$ stands for the power set of $S$. The interior, enclosure, complement and boundary of $S$ are denoted by  $S^{\circ}$, $\bar{S}$, $S^c$ and $\partial S$, respectively. Given a vector $\bm{x}\in \mathbb{R}^n$, $x_i$ denotes the $i$-th coordinate of $\bm{x}$ for $i\in \{1,2,\ldots,n\}$. Also, the  polynomial ring  over variables $\bm{x}$ with coefficients in the  real number field $\mathbb{R}$ is denoted by $\mathbb{R}[\bm{x}]$. $\sum[\bm{x}]$ is used to represent the set of sum-of-squares polynomials over variables $\bm{x}$, i.e.,
$\sum[\bm{x}]=\{p\in \mathbb{R}[\bm{x}]\mid p=\sum_{i=1}^{k} q_i^2,q_i\in \mathbb{R}[\bm{x}],i=1,\ldots,k\}$.

\subsection{Hybrid automata}
HSs of interest in this paper are represented by hybrid automata, defined by \begin{definition}[Hybrid Automaton (HA)]\label{def:HA}
	A HA $\mathcal H$ is a tuple  $(\mathcal Q, \mathcal X, \bm{f}, \mathtt{Init}, \mathtt{Dom}, \mathcal E, \mathcal G, \mathcal R)$, where
	 \vspace*{-2mm}
	\begin{itemize}
		\item  $\mathcal Q = \{q_1, q_2, \dots\}$ is a set of \emph{modes};
		\item $X=\{\mathbf{x}_1,\ldots, \mathbf{x}_n\}$  is a set of \emph{continuous state variables}, which are interpreted over $\mathbb R^n$. Normally, we use 
		 $\mathcal X \subseteq \mathbb R^n $ to denote the continuous state space, and 
		 a (hybrid) state of the system is represented as $(q, \bm{x})\in \mathcal{Q}\times \mathcal{X}$; 
		\item $\mathtt{Init} \subseteq \mathcal Q \times \mathcal X $ is a set of \emph{initial states}; 
		\item $\mathtt{Dom} : \mathcal Q \rightarrow  P(\mathcal X)$  assigns to each  $q\in Q$ a set $\mathtt{Dom}_q$; 
		\item $\bm{f} :\mathcal Q \rightarrow 
		  (\mathcal X \rightarrow \mathbb R^n)$ assigns to each  $q\in Q$ a 
		  \emph{locally Lipschitz continuous vector field} $\textbf{f}_q$ defined on $\mathtt{Dom}_q$; 
		\item $\mathcal E \subseteq \mathcal Q \times \mathcal Q$ is a set of \emph{edges}; 
		\item $\mathcal G: \mathcal E \rightarrow  P(\mathcal X)$ assigns a \emph{guard condition} $\mathcal G_e$ to each edge $e$, such that the discrete jump can happen only if its guard is satisfied; 
		\item $\mathcal R(\cdot, \cdot) : \mathcal E \times \mathcal X \rightarrow  P(\mathcal X)$ assigns a \emph{reset map} to each edge, that relates a state in the pre-mode to a set of states in the post-mode of the edge. 
	\end{itemize}
\end{definition}

Basically, there are two types of evolutions in a HA $\mathcal{H}$, i.e.,  continuous trajectory and discrete jump. A trajectory to  $\bm{f}_q$  with initial state $\bm{x}_0\in \mathtt{Dom}_q$ is a function, denoted by $\bm{\phi}(q, \bm{x}_0, t)$ ($q$ may be omitted if it is clear), of $t$ s.t. $\bm{\phi}(q, \bm{x}_0, 0)=\bm{x}_0$ and its time derivative satisfies  $\frac{\partial \bm{\phi}(q, \bm{x}_0, t)}{\partial t}=\bm{f}_q(\bm{x}_0, t)$. A HA behaves roughly as follows:  it starts with an initial state $(q_0, \bm{x}_0)\in \mathtt{Init}$, and  the continuous state $\bm{x}(t)$  evolves according to  $\bm{f}_{q_0}$ with $\bm{\phi}(q_0, \bm{x}_0, 0)=\bm{x}_0$, 
while the discrete state (mode) $q$ remains unchanged, i.e.,  $q(t)=q_0$, subject to 
$\bm{\phi}(q_0, \bm{x}_0, t)\in \mathtt{Dom}_{q_0}$. If at some time $t_0'$ the continuous state $\bm{x}(t_0')$ meets  the guard $\mathcal{G}_e=\mathcal G(q_0, q_1)$ of some edge $e=(q_0, q_1)\in \mathcal E$, a discrete jump from $q_0$ to $q_1$ 
may happen. Meanwhile, the continuous state is reset to some value according to  $\mathcal R(e, \bm{x})$. After the discrete jump, the continuous evolution resumes and the whole process is repeated. Thus, there is a sequence of time intervals $\bm{\tau} =\{I_i\}_{i=0}^N$, called hybrid time set,  with 
\begin{itemize}
	\item $I_i=[t_i, t_i']$ for all $i<N$;
	\item if $N < \infty$, then $I_N=[t_N, t_N']$ or $I_N=[t_N, t_N')$, where 
	$t_N'$ could be $\infty$; and
	\item  $t_i \leq  t_i' = t_{i+1}$ for all $i$.
\end{itemize}
	A \emph{trace} of a HA $H$ is a sequence of 
	$(q_0,\bm{\phi}(q_0, \bm{x}_{q_0}, \cdot)),\cdots, (q_N,\bm{\phi}(q_N,\bm{x}_{q_N},\cdot))$, shortened as $(\bm{\sigma},\bm{\phi},\bm{\tau})$,   which 
	 satisfies
	\begin{itemize}
		\item[i] Initial condition: $\bm{\phi}(q_0, \bm{x}_{q_0}, 0)= \bm{x}_{q_0}$ with  $(q_0, \bm{x}_{q_0})\in \mathtt{Init}$.
		\item[ii]  Continuous trajectory: for all $i\leq N$,  $\bm{\phi}(q_i, \bm{x}_{q_i}, \cdot):I_i \mapsto \mathbb{R}^n$ is a solution to $\dot{\bm{x}}=\bm{f}(q_i, \bm{x})$, and $\bm{\phi}(q_i,\bm{x}_{q_i},t)\in \mathtt{Dom}_{q_i}$ for any $t\in I_i$.  
		\item[iii]  Discrete jump: for all $i<N$,  $e=(q_i,q_{i+1})\in \mathcal{E}$,  $\bm{\phi}(q_i,\bm{x}_{q_i},t'_i)\in \mathcal{G}_e$ and  
		$\bm{x}_{q_{i+1}} \in  \mathcal{R}(e, \bm{\phi}(q_i,\bm{x}_{q_i},t'_i))$.
	\end{itemize}
A trace with the above form is called \emph{finite} if $I_i$ are {closed} and $N<\infty$; \textit{infinite} if $N =\infty$ or $\sum_{i=1}^N t_i'-t_i =\infty$; \textit{Zeno} if  $N =\infty$ and $\sum_{i=1}^N t_i'-t_i < \infty$; and \textit{maximal} if  it is not a proper  prefix of any other trace of $H$. 

A state $({q},{\bm{x}}) \in \mathcal Q \times \mathcal X$ of $H$ is called \emph{reachable} if there is  a trace $(\bm{\sigma},\bm{\phi},\bm{\tau})$ of $H$ such that $({q},{\bm{x}}) $ is the end state of $(\bm{\sigma},\bm{\phi},\bm{\tau})$. 
 In what follows, we will use  $\mathtt{Reach}_{\mathcal{H}}$ to 
 denote the set of all reachable states of $\mathcal{H}$. 

\subsection{Problem Formulation}
Given a HA $\mathcal{H}$ as Definition \ref{def:HA} and a set of states $\mathcal{S} \subseteq Q\times \mathbb{R}^n$ ( let $\mathcal S_q \, \define \, \{\bm{x} \mid (q, \bm{x}) \in \mathcal S\}$ in what follows), 
we say $\mathcal{H}$ is \emph{safe} w.r.t.  
$\mathcal{S}$, if for any state $(q, \mathbf{x})$ reachable 
in $\mathcal{H}$  $(q, \mathbf{x})\in \mathcal{S}$ holds. 
 
\begin{definition}[Reset Controller Synthesis] \label{def:synthesis}
Given a HA $\mathcal{H}$ as Definition
\ref{def:HA},  we are interested in  the following two types of 
reset controller synthesis problems: 
\begin{itemize}
    \item \textbf{Problem I:} for a given safe set $S$, whether we can find a new $\mathtt{Init}^r$ and $\mathcal{R}^r$ 
such that $\mathcal{H}'= (\mathcal Q,  X, \bm{f}, \mathtt{Init}^r, \mathtt{Dom},$ $\mathcal E, \mathcal G, \mathcal{R}^r)$ is safe with respect to $\mathcal{S}$;
    \item \textbf{Problem II:}    for a given safe set $S$ and a target set $T$, whether we can find a new $\mathtt{Init}^r$ and $\mathcal{R}^r$ 
such that for any $(q,\bm{x})\in \mathtt{Init}^r$ any trace starting with $(q,\bm{x})$ must reach a state in $T$, and the corresponding refined HA is safe  with respect to $\mathcal{S}$ before reaching into $T$.
\end{itemize}
 \end{definition}	
 
Note that the two problems in Definition~\ref{def:synthesis} may 
be solved  by 
synthesizing feedback controllers \cite{tomlin2000,tomlin2000,ames2016control,Paulo0032856,Girard12} or switching logic 
controllers \cite{ZhaoZK13,2009SwitchedSystem,TalyT10} in some cases, but 
may not at all as indicated in Example~\ref{ex:running}. 

\subsection{Transverse set and reach-avoid set}
To address the above two problems, we introduce 
the notions of transverse set and reach-avoid set, and study how to compute them, respectively based on 
\cite{LiuZZ11} and \cite{xue2020}. 

\begin{definition}[Transverse Set \cite{LiuZZ11}] \label{def:tranverse}
Given a vector field $f$  and a semi-algebraic set $S$, the transverse set $\mathtt{trans}^*_{\bm{f}\uparrow S}$ of $f$ over $S$ is defined by 
\begin{equation}
     \mathtt{trans}^*_{\bm{f}\uparrow S } = \{\bm{x}\in \partial S \mid  \forall \epsilon > 0 ~\exists t \in  [0,\epsilon). {\phi}(\bm{x},t) \notin S\}
\end{equation}
\end{definition}
Intuitively, $\bm{x}\in \mathtt{trans}^*_{\bm{f}\uparrow S }$ means that the trajectory ${\phi}(\bm{x},\cdot)$ exits $S$ instantaneously. 
\oomit{See Fig~\ref{fig:DIdiagram}, wherein the dash line is the part of the boundary   not contained in $S$, written as  $\partial^\circ S$, while
the solid line is the part  of  the boundary contained in $S$, written as  $\bar{\partial}S$. It is clear that $x_2$, $x_3$, $x_4$ are in $ \mathtt{trans}^*_{\bm{f}\uparrow S }$, while $x_0$ and $x_1$ are not. }

 As discussed in \cite{LiuZZ11}, $\mathtt{trans}^*_{\bm{f}\uparrow A}$  can be specified in terms of Lie derivatives of   the functions  that define  $S$ along 
 $\bm{f}$. For simplicity, we assume  
   $S \, \define \,  \{\bm{x} \mid {g}( \bm{x}) \leq  0\}$   and   
   $\partial S = \{\bm{x} \mid {g}(\bm{x}) = 0\}$, where 
   ${g}( \bm{x})$ is polynomial. For treating general semi-algebraic sets, please refer to  \cite{LiuZZ11}.
For integers $k\geq 0$, the $k$-th order Lie derivative $\bm{L}^{k}_{f}g(\bm{x})$ of  $g(\bm{x})$ along the vector field $\bm{f}$ is recursively defined by  
\begin{eqnarray}
   \bm{L}^{0}_{f}g(\bm{x}) = g(\bm{x}),  \quad & \cdots, & \quad \bm{L}^{k+1}_{f}g(\bm{x}) = \left\langle f, \frac{ \partial\bm{L}^{k}_{f}g(\bm{x})}{\partial \bm{x}} \right\rangle 
\end{eqnarray}
where $\langle \cdot, \cdot \rangle$ is the inner product of two vectors. 
Let $\gamma_{g,\bm{f}}(\bm{x})=\min\{k\in \mathbb N\mid L^k_{\bm{f}}g(\bm{x})\neq 0\}$,
if such $k$ exists, otherwise $\gamma_{p,\bm{f}}(\bm{x})=\infty$. 
Let $Trans_{\bm{f}\uparrow g} \, \define \, \{\bm{x}\mid \gamma_{g, \bm{f}}(\bm{x}) <\infty \wedge  L^{\gamma_{g,\bm{f}}(\bm{x})}_{\bm{f}}
g(\bm{x}) > 0 \}$. It was  proved in \cite{LiuZZ11} that $\mathtt{Trans}_{\bm{f}\uparrow g}$ is still  a semi-algebraic set for polynomials $\bm{f}$ and $g$. Therefore, we have the following result:
\begin{theorem}
For polynomials $\bm{f}$ and $g$,  $\mathtt{trans}^*_{\bm{f}\uparrow S }=\mathtt{Trans}_{\bm{f}\uparrow g} \cap \{\bm{x} \mid g(\bm{x}) =0\}$ is computable. 
\end{theorem}

In order to reduce reset controller synthesis problems to reach-avoid problems, we generalize the notion of reach-avoid set \footnote{ In  \cite{xue2020}, for given $f$, initial set $\mathcal X_0$, safe region $S$ and target set $\mathtt{TR}\subset S$ with nonempty interior, a reach-avoid set $\mathtt{RA}(S, f, \mathtt{TR})$  is defined as
\[\mathtt{RA}(S, f, \mathtt{TR})= \{\bm{x}\in \mathcal X_0 \mid \exists T. (\phi(\bm{x},T)\in \mathtt{TR} ~ \wedge ~ \forall t\in [0, T]. \phi(\bm{x},T)\in S) \}\]
} in  \cite{xue2020} to the following one.
\begin{definition}[Reach-Avoid Set] \label{def:reach-avoid}
Given a vector field $f$, a bounded safe set $S$  with nonempty interior and a target set $\mathtt{TR}$,  the reach-avoid set $\mathtt{RA}^*(\mathcal X_0\xrightarrow[f]{S} \mathtt{TR})$ is defined as 
\vspace*{-2mm}
\begin{eqnarray}
\mathtt{RA}^*(\mathcal X_0\xrightarrow[f]{S} \mathtt{TR}) &\define &\{\bm{x} \in \mathcal X_0 \mid \exists T\geq 0.~ \forall t\in[0,T). {\phi}(\bm{x},t)\in  S  \nonumber \\  
& &\hspace*{-1cm} \wedge \forall \epsilon>0.\exists t\in [T,T+\epsilon). {\phi}(\bm{x},t)\in  \mathtt{TR}\}.   
\end{eqnarray}
\end{definition}
\vspace*{-2mm}
Intuitively, the set $\mathtt{RA}^*(\mathcal X_0\xrightarrow[f]{S} \mathtt{TR})$ consists of all  states in $\mathcal X_0$ whose trajectories following $f$ eventually enter the target set $\mathtt{TR}$ at or after some instant $T \geq 0$  while staying inside the safe set $S$ before $T$.

Computing  $\mathtt{RA}^*(\mathcal X_0\xrightarrow[f]{S} \mathtt{TR})$ is not easy in general, however, we show that $\mathtt{RA}^*(\mathcal X_0\xrightarrow[f]{S} \mathtt{TR})$ can be approximated  by $\mathtt{RA}(S, f, \mathtt{TR})$ in  \cite{xue2020}.  
We sketch the idea as follows. 
For $\epsilon>0$, we use the $\epsilon$-neighbourhood of  $\mathtt{TR}$ to define a subset  of  $S$ as 
\begin{eqnarray}
\mathcal N_{\epsilon}(\mathtt{TR}) \, \define \, \{\bm{x} \mid \exists \bm{x}' \in \mathtt{TR}. ||\bm{x}-\bm{x}'|| < \epsilon \}, & \quad & 
\mathtt{TR}_{\epsilon}(S)  \, \define  \, \mathcal N_{\epsilon}(\mathtt{TR}) \cap S
\end{eqnarray}

\oomit{
Obviously, 
\begin{lemma}
If $\bar{S} \cap \mathtt{TR} \neq \emptyset$, then $\mathtt{TR}_{\epsilon}(S) \neq \emptyset$ and $\mathtt{TR}_{\epsilon}(S) \subseteq S$. 
\end{lemma} }

\begin{lemma}\label{lem:subcon}
Suppose  $S$ and  $\mathtt{TR}$  are semi-algebraic sets with $S\cap \mathtt{TR}=\emptyset$, then  $\mathtt{RA}^*(\mathcal X_0\xrightarrow[f]{S} \mathtt{TR}) \subseteq \bigcap\limits_{\epsilon >0} \mathtt{RA}(S, f, \mathtt{TR}_{\epsilon}(S))$.
\end{lemma}
\oomit{
\begin{proof}
Suppose  $\bm{x} \in \mathtt{RA}^*(\mathcal X_0\xrightarrow[f]{S} \mathtt{TR})$,  then there is $T\geq 0$ such that (1) ${\phi}(\bm{x},t)\in  S$ for all $t\in[0,T)$; 
(2) for any $\epsilon >0$ there is some $t \in [T, T+\epsilon)$ with  ${\phi}(\bm{x},t )\in  \mathtt{TR}$. 
As $S\cap \mathtt{TR}=\emptyset$, 
 $\phi(\bm{x},T) \in \mathtt{Trans}_{\bm{f}\uparrow S}$. Thus,  
 ${\phi}(\bm{x},T)\in \bar{S} \cap \overline{\mathtt{TR}}$. 
 Therefore, for an $\epsilon>0$, there must be $t < T$ such that ${\phi}(\bm{x},t) \in \mathcal B({\phi}(\bm{x},T), \epsilon/2) =\{\bm{y} \mid ||\bm{y}-{\phi}(\bm{x},T)|| < \epsilon/2\}\subseteq \mathcal N_{\epsilon}(\mathtt{TR})$. Hence, $\bm{x} \in \mathtt{RA}(S, f, \mathtt{TR}_{\epsilon}(S))$. Therefore, $ \mathtt{RA}^*(\mathcal X_0\xrightarrow[f]{S} \mathtt{TR}) \subseteq  \bigcap\limits_{\epsilon >0} \mathtt{RA}(S,f, \mathtt{TR}_{\epsilon}(S))$.
\end{proof}
}

In general, the inverse of Lemma~\ref{lem:subcon} is not true, as it is likely that some trajectories starting from an initial state can arbitrarily approach the target set and  keep safe, but never reach into the target set. Given a safe set $S$, a trajectory ${\phi}(\bm{x},\cdot)$ of a vector field $f$ is called  \textit{infinitely safe}, or \textit{$\infty$-safe}, w.r.t. $S$,  if  ${\phi}(\bm{x},t)\in S$ for all $t\geq 0$.  In what follows, 
$\infty\!\!-\!\!\textit{Safe}(f, \mathcal X_0, S) \, \define \, \{\bm{x}\in  \mathcal X_0 \mid \forall t\geq 0. {\phi}(\bm{x},t)\in S\}$
denotes the set of initial states which trajectories are  $\infty$-safe w.r.t. $S$. 

The following theorem gives a way to approximate $\mathtt{RA}^*( S \xrightarrow[f]{S}  \mathtt{trans}^*_{{f}\uparrow S}) $. 
\begin{theorem}\label{thm:RAPT}
\begin{eqnarray}
\mathtt{RA}^*( S \xrightarrow[f]{S}  \mathtt{trans}^*_{{f}\uparrow S}) 
& = &  \bigcap\limits_{\epsilon >0} \mathtt{RA}(S, f,  (\mathtt{trans}^*_{\bm{f}\uparrow S})_\epsilon(S)) \setminus 
   \infty\!\!-\!\!\textit{Safe}(\bm{f}, S, S). 
\end{eqnarray}
\end{theorem}

\subsection{Synthesizing differential invariants by computing reach-avoid sets}
\begin{definition}[Differential Invariant \cite{zhan2017book}]
\label{def:Invariant}
A set $C$ is a  differential  invariant  of vector field $f$ w.r.t. a set $S$ if for all $\bm{x}\in C$ and $T\geq 0$
    \[(\forall t\in [0,T]. {\phi}(\bm{x},t)\in S) \implies (\forall t\in [0,T]. {\phi}(\bm{x},t) \in C).\]
\end{definition}
Intuitively, if $C \subseteq S$ is a DI  of  $f$ w.r.t. domain $S$, then  the trajectory $\phi(\bm{x}, \cdot)$ from any $\bm{x} \in C$ must stay in $C$ before leaving  the domain. 

Fortunately, Theorem~\ref{thm:RAPT} gives an effective approximation to  differential invariant (DI) of a vector field w.r.t. a given set. Namely, 
\begin{theorem}\label{thm:DI}
For a semi-algebraic set $S$, 
$C \, \define \,  S\setminus \mathtt{RA}^*( S \xrightarrow[f]{S}  \mathtt{trans}^*_{{f}\uparrow S})$ is a DI of  $ f$ w.r.t.  $S$, if $f$ is polynomial.
\end{theorem}

Now, we can use Theorem \ref{thm:DI} to compute DIs of a given HA as follows. Given a $q\in \mathcal Q$, for simplicity, let $ \mathtt{SD}_q =  \mathtt{Dom}_q \cap \mathcal{S}_q$ be the safe domain,   
\begin{equation}\label{eq:dang}
 \mathtt{RM}_q^\star \, \define \,   \mathtt{RA}^*( \mathtt{SD}_q \xrightarrow[\bm{f}_q]{ \mathtt{SD}_q}  \mathtt{trans}^*_{\bm{f}_q\uparrow \mathtt{SD}_q}) \end{equation}
denote the set of states from which the trajectories will 
leave $\mathtt{SD}_q$ eventually, and  let 
\begin{equation}
\mathtt{Dom}^\star_q  \define  \mathtt{SD}_q \setminus \mathtt{RM}_q^\star   \quad  \quad \text{  and   }   \quad  \quad \mathtt{Init}^\star_q  \define  \mathtt{Dom}^\star_q\cap \mathtt{Init}_q \label{eq:CInit}
\end{equation}

By Theorem \ref{thm:DI}, it immediately follows 
\begin{corollary}\label{cor:SDI}
For semi-algebraic sets $\mathtt{Dom}_q$ and $\mathcal S_q$, $\mathtt{Dom}^\star_q$ is a DI of  $\bm{f}_q$ w.r.t.  $\mathtt{SD}_q$.
\end{corollary}

 \section{Reset Controller Synthesis}
\label{sec:reset_map_synthesis}
In this section, we try to solve \textbf{Problem I\&II}. 

\subsection{Reset Controller Synthesis Only with  Safety}
To address \textbf{Problem I}, we need to guarantee that 
in any mode of the refined HS with a synthesized reset controller, any trajectory from its initial set or any reset set associated with a discrete jump to the mode, either stays inside a safe invariant set of the mode, or safely evolves and then  jumps to  a safe state in another mode if the guard of the jump between them is enabled. 
 
So, our solution is implemented by Algorithm \ref{alg:SafeResetSyn}, in which we use $\mathtt{Pre}(q)$ to denote the set of modes that can reach $q$ via one jump and $\mathtt{Post}(q)$ the set of modes reachable from $q$ via one jump. For each mode $q$, we compute  a safe invariant set by removing all states in $q$ from which the trajectory eventually exits the safe domain  $\mathtt{SD}_q $ (line \ref{alg1:line3} of Algorithm \ref{alg:SafeResetSyn});  for each jump $e=(q,p)$ outgoing from $q$, we compute the reach-avoid set with the target set $\mathtt{Dom}_q^c \cap \mathcal{G}_e$ and the safety $\mathtt{SD}_q$, see line \ref{alg1:line4}-\ref{alg1:line6} in Algorithm \ref{alg:SafeResetSyn}. The set computed  by the above two steps is denoted as $\mathtt{Dom}^r_q$. Then, we synthesize a reset map $\mathcal{R}^r$ of each jump $e=(p,q)$ pointing to $q$ as a subset of $\mathtt{Dom}^r_q$ (line \ref{alg1:line7}-\ref{alg1:line9} in Alrorithm \ref{alg:SafeResetSyn}) and the refined  initial set of mode $q$ as the conjunction of the original initial set $\mathtt{Init}_q$ and $\mathtt{Dom}_q^r$ (line \ref{alg1:line10} of Algorithm \ref{alg:SafeResetSyn}).

\begin{algorithm}[t]
	\caption{Reset Control Synthesis Only with Safety}
	\label{alg:SafeResetSyn}
	\begin{algorithmic}[1]
		\REQUIRE  $\mathcal H = (\mathcal Q, \mathcal X, \bm{f}, \mathtt{Init}, \mathtt{Dom}, \mathcal E, \mathcal G, \mathcal R)$ and safe set $\mathcal{S}$ 
		\ENSURE  $\mathcal{H}^r = (\mathcal{Q}, \mathcal X, \bm{f}, \mathtt{Init}^r, \mathtt{Dom}, \mathcal E, \mathcal G, \mathcal{R}^r)$ satisfying $\mathcal{S}$
		\FOR{each $q\in\mathcal{Q}$}\label{alg1:line1}
		    \STATE $\mathtt{SD}_q\gets \mathcal{S}_q \cap \mathtt{Dom}_q$;\label{alg1:line2}
		    \STATE $\mathtt{Dom}^r_q \gets \mathtt{Dom}^\star_q$ computed by  \eqref{eq:CInit};\label{alg1:line3}
		    \FOR{each $p\in\mathtt{Post}(q) $}\label{alg1:line4}
		  \STATE $\mathtt{Dom}^r_q\gets\mathtt{Dom}^r_q\cup\mathtt{RA}^{*}(\mathtt{SD}_q  \xrightarrow[\bm{f}_q]{ \mathtt{SD}_q}  \mathtt{Dom}_q^c \cap \mathcal{G}_e))$;\label{alg1:line5}
		    \ENDFOR\label{alg1:line6}
		    \FOR{each $p\in  \mathtt{Pre}(q) $}\label{alg1:line7}
		            \STATE define $\mathcal{R}^r(e=(p,q),x)\subset \mathtt{Dom}^r_q$;\label{alg1:line8}
		    \ENDFOR\label{alg1:line9}
		    \STATE $\mathtt{Init}^r_q\gets\mathtt{Init}_q\cap\mathtt{Dom}^r_q$;\label{alg1:line10}
		\ENDFOR\label{alg1:line11}
        \RETURN $\mathcal{H}^r = (\mathcal{Q}, \mathcal X, \bm{f}, \mathtt{Init}^r, \mathtt{Dom}, \mathcal E, \mathcal G, \mathcal{R}^r)$;\label{alg1:line12}
	\end{algorithmic}
\end{algorithm}

The correctness of Algorithm~\ref{alg:SafeResetSyn} is guaranteed by the following theorem. 
\begin{theorem}[Correctness]\label{thm:ssf}
\textbf{Problem I} is solvable if and only if $\mathtt{Init}^r$ obtain from Algorithm \ref{alg:SafeResetSyn} is not empty.
\begin{description}
    \item[Soundness] If $\mathtt{Init}^r$ obtained from Algorithm \ref{alg:SafeResetSyn} is not empty, the resulting $\mathcal{H}^r$ solves \textbf{Problem I}.
    \item[Completeness] If \textbf{Problem I} can be solved by some reset controller, $\mathtt{Init}^r$ obtained from Algorithm \ref{alg:SafeResetSyn} is not empty.
\end{description}
\end{theorem}

\oomit{
\begin{remark}
If we do not require the resulted hybrid automaton non-blocking, 
we can revise the above procedure  by replacing $\mathtt{RM}_q^{\star}$ with $\mathtt{RM}_q^{\diamond}$ in line 4 of Algorithm \ref{alg:SafeResetSyn}, and   $ \mathtt{Dom}_q^c \cap \mathcal{G}_e $ with $\mathtt{Dom}_q^c$ in line 6-7. 
\end{remark} }

\subsection{Reset Controller Synthesis with Safety Together with Liveness}
To address \textbf{Problem II}, a natural and simple solution is to compute the reach-avoid set $\mathtt{RA}^*( \mathtt{SD}_q \xrightarrow[\bm{f}_q]{ \mathtt{SD}_q} \mathtt{TR}_q)$, and 
the reach-avoid sets $\mathtt{RA}^*( \mathtt{SD}_q \xrightarrow[\bm{f}_q]{ \mathtt{SD}_q} \mathcal{G}_e \cap \mathtt{Dom}^c_q)$, for each mode $q\in Q$ and the corresponding jumps for each edge  $e=(q,q')\in \mathcal E$, then define a reset controller by setting the initial set to be 
the intersection of the original initial set and the union of 
the above reach-avoid sets, and mapping  any state in the current mode satisfying the guard of a discrete jump to a subset of 
the union of the reach-avoid sets computed as above for the post-mode of the jump. But unfortunately, it does not work in general considering 
the following  three cases.  
\begin{itemize}
    \item Firstly, there may be a path from $q_1$ to $q_2$, $\cdots$, to $q_n$,  with $\mathtt{TR}_{q_n}= \emptyset$, and without any outgoing edge from $q_n$, see any trajectory from $q_0$ to $q_1$ to $q_3$ in Example~\ref{ex:nondeterm}. Thus, we have to redefine the initial set of $q_n$ to be empty, and block the discrete jump from $q_{n-1}$  to $q_n$, and so on.
    \item Secondly, there may be  a loop among modes, say, $q_1, q_2$, $\cdots, q_n, q_1$. Thus, even in each mode $q_i$ (at least some of them), any trajectory from the initial set in the mode can eventually arrive in the target and keeps safe before hitting the target set, but it is also possible that the trajectory keeps evolving along the loop safely forever, see from $q_0, q_1, q_2, q_0$ in Example~\ref{ex:nondeterm}. 
    \item To avoid the above case, we may require reset controller to map a state in $\mathcal{G}_{e_i}$ (let $e_i=(q_i, q_{i+1})$) to the reach-avoid set  $\mathtt{RA}^*( \mathtt{SD}_{q_{i+1}} \xrightarrow[\bm{f}_{q_{i+1}}]{ \mathtt{SD}_{q_{i+1}}} \mathtt{TR}_{q_{i+1}})$ suppose $(q_i, q_{i+1}, \cdots, q_i)$ appears in the trajectory. But possibly the reach-avoid set overlaps (even coincides) with the union of the reach-avoid sets computed from 
      the guards  of the edges outgoing from the current mode, thus there is no such  reset controller that can enforce a trajectory to reach the target in this mode, see $q_2$ and $q_1$ in the loop $q_1, q_2, q_0,q_1$ in Example~\ref{ex:nondeterm}. 
\end{itemize}

{\small 
\begin{figure}[t]
\centering
\scalebox{0.87}{
  \begin{tikzpicture}
  \centering
\tikzstyle{mode} = [circle, minimum size=2.7cm,inner sep=0pt, text centered, draw=black, fill=white!60]
\node (mode1) [mode, xshift=2.6cm] {};
\node (mode4) [mode, xshift=-2.6cm]{};
\node (mode2) [mode, below of=mode1, xshift=-2.6cm, yshift=-3cm]{ };
\node (mode3) [mode, below of=mode1, xshift=2.6cm, yshift=-3cm] {};

\draw [->,thick]  (mode1) to    node[anchor=south, xshift=-0.6cm,  yshift=0.3cm] {$e_0=(q_0, q_1)$} (mode2);
\node [below of=mode1,xshift=-2.4cm,yshift=-0.9cm] {$\mathcal G_{e_0}=[28, +\infty)$};
\node [below of=mode1, xshift=-2.6cm, yshift=-1.4cm] {$\mathcal{R}(e_0,\bm{x})=\bm{x}$};
\draw [->,thick] (mode2) to  node[anchor=south,  yshift=0.0cm] {$e_1=(q_1, q_2)$} (mode3);
\node [below of=mode1, yshift=-3.5cm] {$ \begin{array}{r} 
\mathcal G_{e_1}=[25, 26]\\
\textcolor{red}{\cup[21,23]}
\end{array}$};
\node [below of=mode1, yshift=-4.1cm] {$\mathcal{R}(e_1,\bm{x})=\bm{x}$};
\draw [->,thick] (mode3) to node[anchor=south, xshift=0.6cm, yshift=0.3cm]{$e_2=(q_2, q_0)$} (mode1);
\node [below of=mode1, xshift=2.4cm, yshift=-0.9cm] {$\mathcal G_{e_2}=[27, 29]$};
\node [below of=mode1, xshift=2.8cm, yshift=-1.4cm] {$\mathcal{R}(e_2,\bm{x})=\bm{x}-4$};
\draw [->,thick] (mode2) to node[anchor=south,xshift=-1.4cm,yshift=0.3cm]{$e_3=(q_1,q_3)$} (mode4);
\node [below of=mode4, xshift=0.5cm, yshift=-1.4cm] {$\mathcal{R}(e_3,\bm{x})=\bm{x}$};
\node [below of=mode4, xshift=0.1cm ,yshift=-0.9cm] {$\mathcal G_{e_3}=[28, 30]$};
\node [above of=mode1, yshift=-0.9cm] {$ \begin{array}{c}
    q_0 :  \\
  \dot{x} =1  \\
\mathtt{Init}_0 = [23,25]    \\
\mathtt{Dom}_0 =  (-\infty, 32) \\
\mathtt{TR}_0 = [29, +\infty]  \\
\mathcal S_0  = [23,32]
 \end{array}$};
\node [above of=mode2, yshift=-0.9cm] {$ \begin{array}{c}
    q_1 :  \\
  \dot{x} =  -1  \\
\mathtt{Init}_1 = [26,30]    \\
\mathtt{Dom}_1 = (22,+\infty)     \\
\mathtt{TR}_1 =  [22,24]      \\
\mathcal S_1  =  [22,32]
 \end{array}$};
\node [above of=mode3, yshift=-0.9cm] {$ \begin{array}{c}
    q_2 :  \\
  \dot{x} =  0.1  \\
\mathtt{Init}_2 = [22, 26]    \\
\mathtt{Dom}_2 =  [22, 28)   \\
\mathtt{TR}_2 =   [24,25]  \\
\mathcal S_2  =  [22, 30]
 \end{array}$};
 \node [above of=mode4, yshift=-0.9cm] {$ \begin{array}{c}
    q_3 :  \\
  \dot{x} =  1  \\
\mathtt{Init}_3 = [22, 26]    \\
\mathtt{Dom}_3 =  [22, 28)   \\
\mathtt{TR}_3 =   \emptyset  \\
\mathcal S_3  =  [22, 30]
 \end{array}$};
\end{tikzpicture}}
\caption{ An HA with safety and liveness: for simplicity, $\mathtt{Init}_{q_0}$ is shorten as $\mathtt{Init}_0$,  similarly to others. }
\label{fig:livenessexp}
\end{figure}
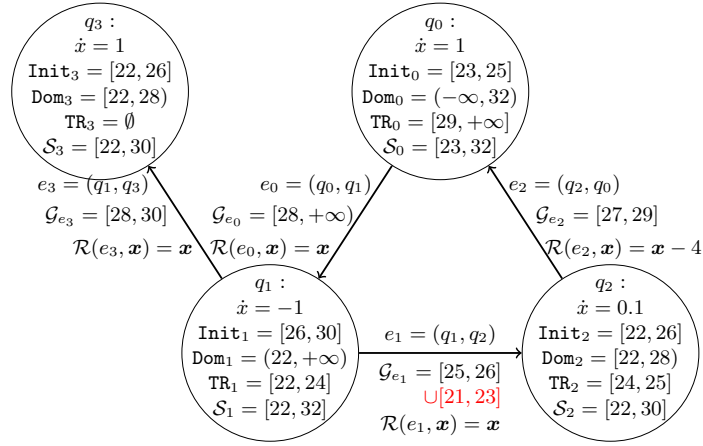 }

The first two cases  can be handled similarly. Considering the second case,  the basic idea of our solution is to define proper reset maps to ``break down'' such kind of loops. In details,  suppose $(q_0,q_1,\cdots,q_k,q_0)$ is a loop, then 
we start from $q_{i-1}$ for some $0< i \leq  k$ and let  
\begin{equation}\label{eq:lresetmap}
\mathtt{ST}_{q_{i}} \define \mathtt{RA}^*( \mathtt{SD}_{q_{i}} \xrightarrow[\bm{f}_{q_{i}}]{ \mathtt{SD}_{q_{i}} ~ \setminus ~ \bigcup\limits_{e'=(q',q_{i})\in \mathcal E}\mathcal{G}_{e'} } \mathtt{TR}_{q_{i}}).  \end{equation}
 If $\mathtt{ST}_{q_{i}}\neq \emptyset$, that means 
 any trajectory from $\mathtt{ST}_{q_{i}}$ cannot  
 satisfy the guard of any discrete jump out of $q_i$ and such trajectories 
 do exist, then  we set   $\mathcal{R}^r(e,\bm{x}) \subseteq \mathtt{ST}_{q_{i}}$. Thus,  any  trajectory starting from $q_{i}$ after the jump $(q_{i-1},q_i)$ can only reach to   the target set in $q_{i}$, i.e., $\mathtt{TR}_{i}$,   therefore the loop is blocked.  Otherwise, we 
 go to $q_{i+1}$ and repeat the above procedure until back to $q_i$ again. 
 If back to $q_i$, that implies that no reset maps can block the loop. 
 Therefore, we have to remove the whole loop by setting the initial set in each mode to be empty and blocking any discrete jumps to the loop. 

Let's consider the following example. 
\begin{example}\label{ex:nondeterm}
Suppose an HA consisting of $\mathcal Q=\{q_0,q_1,q_2,q_3\}$, 
 $\mathcal{E}=\{e_0,e_1,e_2, e_3\}$, $\mathcal X= \mathbb R$, its vector fields, initial sets, domains, guard conditions, reset maps, target sets, and safe regions  given as in Fig. \ref{fig:livenessexp}, in which $(\mathcal{Q},\mathcal{E})$ forms a directed graph. 
According to the approach given above,  we have 
$ \mathtt{RA}^{*}( \mathtt{SD}_{q_0}  \xrightarrow[\bm{f}_{q_0}]{ \mathtt{SD}_{q_0}}  \mathtt{TR}_{q_0})  =[23,32]$,  
 $\mathtt{RA}^{*}( \mathtt{SD}_{q_1}  \xrightarrow[\bm{f}_{q_1}]{ \mathtt{SD}_{q_1}}  \mathtt{TR}_{q_1}) = (22,32]$,  $\mathtt{RA}^{*}( \mathtt{SD}_{q_2}  \xrightarrow[\bm{f}_{q_2}]{ \mathtt{SD}_{q_2}}  \mathtt{TR}_{q_2})  =[22, 25]$, and  $\mathtt{RA}^{*}( \mathtt{SD}_{q_3}  \xrightarrow[\bm{f}_{q_3}]{ \mathtt{SD}_{q_3}}  \mathtt{TR}_{q_3})  =\emptyset$. Also, we have  $\mathtt{RA}^{*}( \mathtt{SD}_{q_0} \xrightarrow[\bm{f}_{q_0}]{ \mathtt{SD}_{q_0}}   \mathcal{G}_{e_0} \cap \mathtt{Dom}^c_{q_0})=[23,32]$,   
 $\mathtt{RA}^{*}( \mathtt{SD}_{q_1} \xrightarrow[\bm{f}_{q_1}]{ \mathtt{SD}_{q_1}}   \mathcal{G}_{e_1} \cap \mathtt{Dom}^c_{q_1})=[25,32]$, 
  $\mathtt{RA}^{*}( \mathtt{SD}_{q_2} \xrightarrow[\bm{f}_{q_2}]{ \mathtt{SD}_{q_2}}   \mathcal{G}_{e_2} \cap \mathtt{Dom}^c_{q_2})=[22, 28]$, 
  and $\mathtt{RA}^{*}( \mathtt{SD}_{q_1} \xrightarrow[\bm{f}_{q_1}]{ \mathtt{SD}_{q_1}}   \mathcal{G}_{e_3} \cap \mathtt{Dom}^c_{q_1})=[28, 32]$.

Clearly, any trajectory starting from $q_3$ cannot reach the target set. Therefore, we have to set the initial set of $q_3$ to be empty. Meanwhile, we also need to block any jump from $q_1$ to $q_3$. To the end, we need to reset the initial set of $q_1$ to be $[26,28)$ and redefine the reset map for $e_0$ such that any state in $q_0$ is reset a value less than $28$. 

Moreover, for any trajectory starting from $\mathtt{Init}_{i}$, $i=0,1,2$, it may reach to the target in every mode, and also may never reach to it if appropriate jump points are selected, for example, the trace $(q_0,[0,4),[24, 28))$, $(q_1,[4,6),[28,26))$, $(q_2, [6,8), [26,28))$, $(q_0$, $[8,10), [24, 28))$, $(q_1, [10,12), [28,26))$, $(q_2, [12, 14), [26,28)), \cdots$. 

Even worse, we cannot find reset maps for $e_2$ and $e_1$ so that any loops like above can be blocked at $q_0$ and $q_2$  to guarantee that the target set is reachable,  because 
{\small  \begin{eqnarray} 
 \mathtt{RA}^{*}( \mathtt{SD}_{q_0}  \xrightarrow[\bm{f}_{q_0}]{ \mathtt{SD}_{q_0}}  \mathtt{TR}_{q_0}) \setminus \mathtt{RA}^{*}( \mathtt{SD}_{q_0} \xrightarrow[\bm{f}_{q_0}]{ \mathtt{SD}_{q_0}}   \mathcal{G}_{e_0} \cap \mathtt{Dom}^c_{q_0}) &= & \emptyset, \\[-2mm] 
  \mathtt{RA}^{*}( \mathtt{SD}_{q_2}  \xrightarrow[\bm{f}_{q_2}]{ \mathtt{SD}_{q_2}}  \mathtt{TR}_{q_2}) \setminus \mathtt{RA}^{*}( \mathtt{SD}_{q_2} \xrightarrow[\bm{f}_{q_2}]{ \mathtt{SD}_{q_2}}   \mathcal{G}_{e_2} \cap \mathtt{Dom}^c_{q_2}) &= & \emptyset.
 \end{eqnarray} }
However, these loops can be blocked in $q_1$ by redefining a reset map  $\mathcal{R}^r(e_0, \bm{x})$ to a subset of $(22,23)$ as 
{\small \begin{eqnarray} \hspace*{-5mm}
 \mathtt{RA}^{*}( \mathtt{SD}_{q_1}  \xrightarrow[\bm{f}_{q_1}]{ \mathtt{SD}_{q_1}}  \mathtt{TR}_{q_1}) \setminus \mathtt{RA}^{*}( \mathtt{SD}_{q_1} \xrightarrow[\bm{f}_{q_1}]{ \mathtt{SD}_{q_1}}   \mathcal{G}_{e_1} \cap \mathtt{Dom}^c_{q_1}) &= & [22,25).  \label{eq:q1}
 \end{eqnarray} }
What's more, if we revise the guard of $e_1$ by cojoining $[21,23]$ (see the red part of $\mathcal{G}_{e_1}$ in Fig.~\ref{fig:livenessexp}), then the left side of \eqref{eq:q1} will become empty, thus   no reset maps exist that can guarantee safety together liveness any more in this example. 
\end{example}

{\small 
\begin{breakablealgorithm}
	\caption{Reset Control Synthesis With Safety Together with Liveness}
	\label{alg:LivResetSyn}
	\begin{algorithmic}[1]
		\REQUIRE  $\mathcal H = (\mathcal Q, \mathcal X, \bm{f}, \mathtt{Init}, \mathtt{Dom}, \mathcal E, \mathcal G, \mathcal R)$ , safe set $\mathcal{S}$ and target set $\mathtt{TR}$
		\ENSURE  $\mathcal{H}^r = (\mathcal{Q}, \mathcal X, \bm{f}, \mathtt{Init}^r, \mathtt{Dom}, \mathcal E, \mathcal G, \mathcal{R}^r)$ that can guarantee that all trajectories can reach to $\mathtt{TR}$  and satisfy $\mathcal{S}$ before reaching $\mathtt{TR}$, or "No Such Reset Controllers Exist"
		\FOR{each $q\in\mathcal{Q}$} \label{alg3:line1}
		    \STATE $\mathtt{SD}_q\gets \mathcal{S}_q \cap \mathtt{Dom}_q$; \label{alg3:line2}
		     \STATE $\mathtt{Dom}^r_q \gets \mathtt{RA}^{*}(\mathtt{SD}_q   \xrightarrow[\bm{f}_q]{ \mathtt{SD}_q} \mathtt{TR}_q)$; \label{alg3:line3}
		   \STATE \textbf{for} each $p\in\mathtt{Post}(q)$ \textbf{do}
             $\mathtt{Dom}^r_q\gets\mathtt{Dom}^r_q\cup\mathtt{RA}^{*}(\mathtt{SD}_q   \xrightarrow[\bm{f}_q]{ \mathtt{SD}_q} 
		         \mathtt{Dom}^c_q \cap  \mathcal{G}_{e=(q,p)}))$ \textbf{end for};\label{alg3:line5}
		    \STATE $\mathtt{Init}^r_q\gets\mathtt{Init}_q\cap\mathtt{Dom}^r_q$, \,
		     $\mathtt{ST}_q\gets\mathtt{ST}_q$ computed by \eqref{eq:lresetmap}; \label{alg3:line8}
		\ENDFOR \label{alg3:line9}
		\STATE \textbf{for} each $q$ with $\mathtt{Init}^r_q\neq\emptyset$ \textbf{do} 
             $\textit{Refining\_Dom}(q)$ \textbf{end for} ;\label{alg3:line11}
        \STATE \textbf{for} each $q\in\mathcal{Q}$ \textbf{do}
             $\mathtt{Init}^r_q\gets\mathtt{Init}_q^r\cap\mathtt{Dom}^r_q$ \textbf{end for};\label{alg3:line14}
	    \FOR{each $e=(p,q)\in\mathcal{E}$} \label{alg3:line17}
	        \STATE $\mathcal{R}^r(e,x)\subseteq\mathtt{Dom}^r_q$ \textbf{if} $\mathcal{R}^r(e,x)$ has not been refined in Algorithm \ref{alg:DomRef};\label{alg3:line18}
	    \ENDFOR \label{alg3:line22}
	    \IF{ $\mathtt{Init}^r = \cup_{q\in Q}\mathtt{Init}^r_q\neq \emptyset$} \label{alg3:line23}
	        \RETURN $\mathcal{H}^r = (\mathcal{Q}, \mathcal X, \bm{f}, \mathtt{Init}^r, \mathtt{Dom}, \mathcal E, \mathcal G, \mathcal{R}^r)$; \label{alg3:line24}
	    \ELSE \label{alg3:line25}
	        \RETURN ``No Such Reset Controller Exist"; \label{alg3:line26}
	    \ENDIF \label{alg3:line27}
	\end{algorithmic}
\end{breakablealgorithm}
}

{\small 
\begin{breakablealgorithm}
	\caption{$\textit{Refining\_Dom}(q)$}
	\label{alg:DomRef}
	\begin{algorithmic}[1]
		\STATE $\textit{MS}\gets\emptyset$; \label{alg2:line1}
		//MS is a stack to store modes to be visited \\
		// \textit{Path} is an array with length $|Q|+1$, stores the path under consideration
		    \STATE  $\textit{Tag}[-1] \gets \textit{true}$ ;
		          \textbf{for} 
		      $j=0$ \textbf{to} $|Q|$ \textbf{do}   $\textit{Tag}[i] \gets \textit{false}$ \textbf{end for}; \label{alg2:line2} \\
		      //$\textit{Tag}[j]$ indicates whether the mode 
		         $\textit{Path}[j]$ in the considered path 
		          has an unexplored child stored in $\textit{MS}$
		    \STATE \textbf{push}$(q,\textit{MS})$; $j\gets 0$; \label{alg2:line3}
		    \WHILE{$\textit{MS}\neq\emptyset$} \label{alg2:line4}
                \STATE $q\gets \textbf{pop}(\textit{MS})$; $\textit{Path}[j]\gets q$; \label{alg2:line5}
                \IF{$\mathtt{Post}(q)=\emptyset$} \label{alg2:line6}
                    \STATE $i\gets j$; $j\gets\sup_{k<j}\{k\mid \textit{Tag}[k]\}+1$; \label{alg2:line7}
                    \WHILE{$i \ge j \wedge i>0$} \label{alg2:line8}
                        \IF{$\mathtt{ST}_{Path[i]}\neq\emptyset$} \label{alg2:line9}
                            \STATE $\mathcal{R}^r(e=(\textit{Path}[i-1],\textit{Path}[i]),\bm{x})\subseteq ST_{\textit{Path}[i]}$; \label{alg2:line10}
                            \STATE Break; \label{alg2:line11}
                        \ELSE \label{alg2:line12}
                            \STATE 
                    $\mathtt{Dom}^r_{\textit{Path}[i-1]}\gets\mathtt{Dom}^r_{\textit{Path}[i-1]}\setminus\mathtt{RA}^{*}(\mathtt{SD}_{\textit{Path}[i-1]}\xrightarrow[\bm{f}_{\textit{Path}[i-1]}]{ \mathtt{SD}_{\textit{Path}[i]}}\mathcal{G}_{e=(\textit{Path}[i-1],\textit{Path}[i])})$; \label{alg2:line13}
                            \STATE $i\gets i-1$; \label{alg2:line14}
                        \ENDIF \label{alg2:line15}
                    \ENDWHILE \label{alg2:line16}
                \ELSE \label{alg2:line17}
                    \IF{$q=\textit{Path}[i]$ for some $0\le i \le j$ } \label{alg2:line18}
                        \STATE $l\gets j$;  $j\gets\sup_{k<j}\{k\mid \textit{Tag}[k]= \textit{true} \}+1$; \label{alg2:line19}
                        \WHILE{$k\ge j\wedge l>i $} \label{alg2:line20}
                        \IF{$ST_{\textit{Path}[k]}\neq\emptyset$}\label{alg2:line21}
                                \STATE $\mathcal{R}^r((\textit{Path}[l-1],\textit{Path}[l]),x)\subseteq ST_{\textit{Path}[l]}$; \label{alg2:line22}
                                \STATE Break; \label{alg2:line23}
                            \ELSE \label{alg2:line24}
                                \STATE 
                     $\mathtt{Dom}^r_{\textit{Path}[l-1]}\gets\mathtt{Dom}^r_{\textit{Path}[l-1]}\setminus\mathtt{RA}^{*}(\mathtt{SD}_{\textit{Path}[l-1]}\!\xrightarrow[\bm{f}_{\textit{Path}[l-1]}]{ \mathtt{SD}_{\textit{Path}[l]}}\mathcal{G}_{e=(\textit{Path}[l-1],\textit{Path}[l])})$; \label{alg2:line25}
                                \STATE $l \gets l-1$; \label{alg2:line26}
                            \ENDIF \label{alg2:line27}
                        \ENDWHILE \label{alg2:line28}
                    \ELSE \label{alg2:line29}
                        \STATE \textbf{for} {each $p\in\mathtt{Post}(q)$}
                            \textbf{do} \textbf{push}$(p,\textit{MS})$ \textbf{end for}; \label{alg2:line30}
                        \STATE \textbf{if} 
                        {$|\mathtt{Post}(q)|>1$} \textbf{then} 
                         $\textit{Tag}[j]\gets \textit{true}$ \textbf{end if}; \label{alg2:line31}
                        \STATE $j\gets j+1$; \label{alg2:line32}
                    \ENDIF \label{alg2:line33}
                \ENDIF \label{alg2:line34}
            \ENDWHILE \label{alg2:line35}
	\end{algorithmic}
\end{breakablealgorithm}
}

We implement the above idea in Algorithm \ref{alg:LivResetSyn}. 
In line \ref{alg3:line1}-\ref{alg3:line9} of Algorithm \ref{alg:LivResetSyn}, we compute $\mathtt{Init}^r$  in a similar way as in Algorithm \ref{alg:SafeResetSyn}. The only differences lies in that at line~\ref{alg3:line3}  we initialize $\mathtt{Dom}^r_q$ with the reach-avoid set of the target set $\mathtt{TR}_q$ w.r.t. $\mathtt{SD}_q$ rather than  the maximal DI contained in $\mathtt{SD}_q$ computed by \eqref{eq:CInit}.  Additionally, at line~\ref{alg3:line8}, we compute the part of the reach-avoid set of $\mathtt{TR}_q$ that is disjoint with the guard of any discrete jump out of $q$, which will be used in the procedure $\textit{Refining\_Dom}$. At Line \ref{alg3:line11}, for each mode $q$ with $\mathtt{Init}_q^r\neq \emptyset$, $\textit{Refining\_Dom}$ is invoked to check whether  any path starting from  $q$ with length (here meaning the number of discrete modes occurring in it)  no more than $|Q|+1$ is one of the three aforementioned cases in a depth first manner by re-calculating  the corresponding domain constraints on the path, which is used to re-define the initial set of each mode (line \ref{alg3:line14}), and the reset map for each discrete jump (line \ref{alg3:line17}-\ref{alg3:line22}). Based on which, either a reset controller solving the problem is synthesized  (line~\ref{alg3:line23}-\ref{alg3:line24}), or ``no such reset controller exist"  is reported (line \ref{alg3:line25}-\ref{alg3:line26}). 

Algorithm~\ref{alg:DomRef} applies  depth-first searching all possible paths from $q$ with at most $|Q|+1$ modes. To the end, the stack $\textit{MS}$ is used to store modes to be visited; $\textit{Path}$, an array of length $|Q|+1$, to store the current path under consideration; $\textit{Tag}$, an array of length  $|Q|+2$, to indicate whether the $i$-th mode in $\textit{Path}$ has a child that is not explored and deposited in $\textit{MS}$. For easing treating, we always let $\textit{Tag}[-1]= \textit{true}$. Line~\ref{alg2:line1}-\ref{alg2:line3} do the standard initialization. At line~\ref{alg2:line5}, we pop an unexplored  mode from $\textit{MS}$, and deposit it in $\textit{Path}[j]$, as there is a discrete jump $e=(\textit{Path}[j-1],\textit{Path}[j])$ if $j>0$. Then, we test whether $q$ has outgoing discrete jump(s) (line~\ref{alg2:line7}). If no, we should go back to the nearest mode (computed at line~\ref{alg2:line7})  from $q$ which still has outgoing discrete jump(s) unexplored (line~\ref{alg2:line8}-\ref{alg2:line16}). Meanwhile, in each backtracking step, we need to distinguish two cases depending on whether $\mathtt{ST}_{\textit{Path}[i]}=\emptyset$. If $\mathtt{ST}_{\textit{Path}_{i}}\neq \emptyset$, which means that any trajectory from $\mathtt{ST}_{\textit{Path}[i]}$ can safely reach to the target set in this mode, then we just need to set the reset map for $e=(\textit{Path}[i-1], \textit{Path}[i]$ to be a subset of $\mathtt{ST}_{\textit{Path}[i]}$, exit the loop and go back to the nearest mode from the end of the path with unexplored jump(s) stored in $\mathit{MS}$ (line~\ref{alg2:line9}-\ref{alg2:line11}); otherwise, we have to strengthen the domain of $\textit{Path}[i-1]$ that can block the jump from $\textit{Path}[i-1]$ to $\textit{Path}[i]$ (line~\ref{alg2:line12}-\ref{alg2:line14}). If $q$ has occurred in the path, i.e., there is a loop from $q$ to $q$ in the path, the treatment is quite similar to the above case (line~\ref{alg2:line18}-\ref{alg2:line28}); otherwise, we just push each $p$ in $\mathtt{Post}(q)$ into $\textit{MS}$ (line~\ref{alg2:line30}), then tag that the current mode has unexplored  discrete jump(s)  if $|\mathtt{Post}(q)|>1$  (line~\ref{alg2:line31}), and increase $j$ by $1$.

As the termination of Algorithm~\ref{alg:DomRef} is obvious, so  
\begin{proposition}
Algorithm \ref{alg:LivResetSyn} always terminates at finitely many steps.
\end{proposition}

Note that if the $\mathtt{Dom}^r_q$  defined by line 8 of Algorithm \ref{alg:LivResetSyn} is empty then  $\mathtt{Init}^r_q$ must be empty. So, we could regard that the initial states of $q$ satisfying the target reachiability is vacuously true. Without loss of generality, for each $q\in \mathcal Q$ we assume  $\mathtt{Dom}^r_q$ is \emph{nonempty} in the following discussion.
\begin{theorem}[Correctness]\label{thm:lsf}
\textbf{Problem II} is solvable if and only if $\mathtt{Init}^r$ obtain from Algorithm \ref{alg:LivResetSyn} is not empty. 
\qquad
\begin{description}
    \item[Soundness] Our approach is sound, that is, any reset controller synthesized by the above approach does solve 
    \textbf{Problem II};
    \item[Completeness] Our approach is also complete, that is, 
      if \textbf{Problem II} can be solved by some reset controller, the above approach does synthesize such one. 
\end{description}
\end{theorem}

\section{An SDP Approach to Computing Reach-Avoid Sets}
\label{sec:SDPavoid}

The concept of reach-avoid set $\mathtt{RA}^*(\mathcal X_0\xrightarrow[\bm{f}]{S} \mathtt{TR})$ plays a crucial role in Algorithms \ref{alg:SafeResetSyn}\& \ref{alg:LivResetSyn}\&\ref{alg:DomRef}. 
In order to find an inner-approximate $\mathtt{RA}^*(\mathcal X_0\xrightarrow[\bm{f}]{\mathcal S} \mathtt{TR})$, our solution is to  compute a 
conservative inner-approximation to  $\mathtt{RA}^*(\mathcal{S}\xrightarrow[\bm{f}]{\mathcal{S}} \mathtt{TR})$ first,  then 
show the intersection of the inner-approximation and the initial set 
provides a tighter inner-approximation of  $\mathtt{RA}^*(\mathcal X_0\xrightarrow[\bm{f}]{\mathcal S} \mathtt{TR})$.\oomit{ Even though each $\mathtt{Dom}_q$, $\mathcal G_e$, and  $\mathcal S_q$ are defined by a single polynomial inequality constraint, the target sets in Algorithms \ref{alg:SafeResetSyn}\&\ref{alg:LivResetSyn}\&\ref{alg:DomRef} are defined by multiple polynomial constraints.} To the end, we inner-approximate the reach-avoid set in the first step  by extending  Corollary 3 in \cite{xue2020} to polynomial hybrid systems as follows.
\begin{theorem}	\label{thm:IAP}
Given a polynomial vector field $\bm{f}(\bm{x})$, and 
semi-algebraic sets $\mathcal X_0$, $\mathcal S$ which is 
bounded, and $\mathtt{TR}$ which is open and 
represented by $\{\bm{x}\mid \wedge_{k=1}^K \gamma_k(\bm{x})<0\}$, if there exist polynomials $\theta(\bm{x}) \in \mathbb{R}[\bm{x}]$ and $\psi(\bm{x}) \in \mathbb{R}[\bm{x}]$ such that 
	\begin{align}
	-\left\langle \frac{\partial \theta(\bm{x})}{\partial\bm{x}}, \bm{f}(\bm{x}) \right\rangle \geq 0,\ & \forall \bm{x} \in \mathcal{S} \setminus  \mathtt{TR}  \label{eq:racon1}\\
	\theta(\bm{x}) \geq  \max\limits_{k\in \{1,\cdots,K\}}\gamma_k(\bm{x}) + \left\langle\frac{\partial \psi(\bm{x})}{\partial\bm{x}}, \bm{f}(\bm{x})\right\rangle,\ & \forall \bm{x} \in\mathcal{S} \setminus  \mathtt{TR}  \label{eq:racon2} 
	\\
	\theta(\bm{x}) \geq 0,\ & \forall \bm{x} \in  \partial \mathcal{S} \label{eq:racon3}
	\end{align}
	hold, then $\{\bm{x} \in  \mathcal{S} \mid \theta(\bm{x})<0\}$ is an inner-approximation of $\mathtt{RA}^*(\mathcal S \xrightarrow[\bm{f}]{\mathcal S} \mathtt{TR})$. Furthermore, $ \{\bm{x} \in \mathcal X_0 \mid \theta(\bm{x})<0\}$ is a tighter  inner-approximation of   $\mathtt{RA}^*(\mathcal X_0\xrightarrow[\bm{f}]{\mathcal S} \mathtt{TR})$.
\end{theorem}

\begin{algorithm}[t]
	\caption{Inner-Approximation of Reach-Avoid Sets}
	\label{alg:reachavoid}
	\begin{algorithmic}[1]
		\REQUIRE Initial set $\mathcal X_0$, safety set $\mathcal S=\{\bm{x} \mid \wedge_{j=1}^J \zeta_j(\bm{x}) < 0\}$, target set  $\mathtt{TR}=\{\bm{x}\mid \wedge_{k=1}^K \gamma_k(\bm{x})<0\}$, polynomial templates $\theta(\bm{\alpha}, \bm{x})\in \mathbb{R}[\bm{\alpha}, \bm{x}]$, $\psi(\bm{\beta}, \bm{x})\in \mathbb{R}[\bm{\beta}, \bm{x}]$, $s_{i,j}(\bm{\alpha_{i,j}}, \bm{x})\in \sum[\bm{\alpha_{i,j}}, \bm{x}]$, $i=1,2,3$, and $j =1,2, \cdots, J$, and $h_{i,k}(\bm{\beta_{i,k}},\bm{x})\in \sum[\bm{\beta_{i,k}}, \bm{x}]$, $i=1,2$ and $k =1,2, \cdots, K$, of appropriate degrees, with coefficient parameters $\bm{\alpha}, \bm{\beta}, \bm{\alpha_{i,j}}$, and $\bm{\beta_{i,k}}$. For simplicity, let $\theta, \psi, s_{i,j}$, and $h_{i,k}$ denote these polynomials.
		\ENSURE An inner-approximation of $\mathtt{RA}^*(\mathcal X_0\xrightarrow[\bm{f}]{\mathcal S} \mathtt{TR})$
		\STATE Solve the equation:
		\begin{equation}
		\label{sos}
		\begin{split}
		&\min \quad \bm{\alpha} \cdot \bm{w}\\
		&\text{s.t.}\left\{\begin{array}{ll}
		&\bigwedge_{k=1}^{K}-\left\langle\frac{\partial \theta}{\partial \bm{x}},\bm{f}(\bm{x})\right\rangle + \sum_{j=1}^{J} s_{1,j}\zeta_j(\bm{x}) -h_{1,k}\gamma_k(\bm{x}) \in \sum[\bm{x}];\\
		&\bigwedge_{k=1}^{K} \theta-\left\langle \frac{\partial \psi}{\partial \bm{x}}, \bm{f}(\bm{x})\right\rangle + \sum_{j=1}^{J} s_{2,j}\zeta_j(\bm{x}) -  h_{2,k}\gamma_k(\bm{x}) \in \sum[\bm{x}];\\
		&\theta - \sum_{j=1}^{J} s_{3,j}\zeta_j(\bm{x}) \in \sum[\bm{x}]
		\end{array}\right.
		\end{split}
		\end{equation}
		\IF{\eqref{sos} is solved successfully with optimal argument  $\bm{\alpha}^*, \bm{\beta}^*, \bm{\alpha}^*_{i,j}$, and $\bm{\beta}^*_{i,k}$}
		\RETURN $\{\bm{x}\in \mathcal X_0 \mid \theta(\bm{\alpha}^*, \bm{x})<0 \}$;
		\ENDIF
	\end{algorithmic}
\end{algorithm}

We compute $\{\bm{x} \in  \mathcal{S} \mid \theta(\bm{x})<0\}$ in  Theorem \ref{thm:IAP} by solving constraints \eqref{eq:racon1}-\eqref{eq:racon3} for a given template $\theta(\bm{\alpha}, \bm{x})$, 
which can be encoded as 
semi-definite constraints using the sum-of-squares decomposition for multivariate polynomials. As the initial and safe sets in Algorithms \ref{alg:SafeResetSyn} and \ref{alg:LivResetSyn} are defined by multiple polynomial constraints, we focus on initial sets of form $\mathcal X_0=\{\bm{x}\mid \wedge_{i=1}^I \eta_i(\bm{x}) \geq 0\}$ and safe sets of form $\mathcal S=\{\bm{x} \mid \wedge_{j=1}^J \zeta_j(\bm{x}) < 0\}$.  We set the objective function to be 
$\bm{\alpha} \cdot \bm{w}=\int_{\mathcal S}\theta(\bm{\alpha}, \bm{x})d\bm{x}$,
where $\bm{w}$ is the constant vector computed by integrating the monomials in $\theta(\bm{\alpha}, \bm{x})\in \mathbb{R}[\bm{\alpha}, \bm{x}]$ over $\mathcal S$, $\alpha$ is the vector composed of unknown coefficients,  $\psi(\bm{\beta}, \bm{x})\in \mathbb{R}[\bm{\beta}, \bm{x}]$, and $s_{i,j}(\bm{\alpha}_{i,j}, \bm{x})\in \sum[\bm{\alpha}_{i,j}, \bm{x}]$, $i=1,2,3$ and $j =1, 2, \cdots, J$, and $h_{i,k}(\bm{\beta}_{i,k}, \bm{x})\in \sum[\bm{\beta}_{i, k}, \bm{x}]$, $i=1, 2$ and $k =1, 2, \cdots, K$, where $\bm{\beta}$, $\bm{\alpha}_{i, j}$, and $\bm{\beta}_{i, k}$ are indeterminate coefficients. If the solution to SDP \eqref{sos} is feasible, then Algorithm \ref{alg:reachavoid} returns an inner-approximation to $\mathtt{RA}^*(\mathcal X_0\xrightarrow[\bm{f}]{\mathcal S} \mathtt{TR})$.

\begin{theorem}	\label{inner0}
Let $\theta(\bm{\alpha}^*, \bm{x})$ be a solution to \eqref{sos}, then Algorithm \ref{alg:reachavoid} returns an inner-approximation to  $\mathtt{RA}^*(\mathcal X_0\xrightarrow[\bm{f}]{\mathcal S} \mathtt{TR})$.
\end{theorem}

How to predefine  templates plays a significant role in Algorithm~\ref{alg:reachavoid}. If the semi-definition program \eqref{sos} is not solved successfully, one ordinary routine is to use polynomial templates with higher degrees in \eqref{sos}. Although computational cost becomes prohibitive as either the dimension of the vector field $\dot{\bm{x}}=\bm{f}(\bm{x})$ or the polynomial degree of $\theta(\bm{\alpha}, \bm{x})$ and/or $\psi(\bm{\beta}, \bm{x})$ increases, at least with the standard approach to the sum-of-squares optimization wherein generic semi-definite programs are solved by second-order symmetric interior-point algorithms. Larger problems may be tackled using specialized non-symmetric interior-point \cite{papp2019sum} or first-order algorithms \cite{zheng2018fast}.

  \section{Case Study}
\label{sec:ier}

The above algorithms are implemented  using 
the sum-of-squares module in YALMIP \cite{yalmip}, which is a toolbox for modeling and optimization in MATLAB, and the semi-definite programming solver  MOSEK \cite{mosek}. 
We will illustrate  the efficiency and effectiveness of our approach 
by applying the implementation to  van der Pol oscillators, 
aircraft conflict resolution and CWH equations, 
where CWH equations comes from the practice of space. 

\oomit{
are used to  illustrate the efficiency of our algorithms: 
\begin{enumerate}
	\item Van der Pol oscillators, which is modelled as 
	an HA (two modes) with nonlinear dynamics. We will synthesize  a safe reset controller for it.
	\item Aircraft conflict resolution, which 
	 is modelled as an HA with two modes. We will synthesize
	 a reset controller to guarantee  safety and  liveness for it.
	\item CWH equations that describes a simplified model of orbital relative motion of two spacecrafts, which is modelled 
	as an HA with three modes.  For which we will synthesize a reset controller 
	to guarantee safety together with liveness. 
\end{enumerate} }

All experiments are conducted on a Windows 11 PC with Intel(R) Core(TM) i$5-8265$U CPU @$1.6$GHz and $8.00$ GB RAM. The running time of our case studies is as \cref{tab:time_consuming}.

\subsection{Van der Pol oscillator}
Van der Pol oscillator is an oscillator with nonlinear damping governed by the second-order differential equation, which is in widely use in electronic engineering. Consider that two classical van der Pol oscillators are combined by discrete jumps, 
which can be represented by a HA as in Fig.\ref{fig:ex1}. The vector fields for the two modes are
\begin{equation*}
\begin{aligned}
      q_1:\left\{ \begin{array}{lr}
     \dot{x}_1  = -2 x_2, &\\
   \dot{x}_2 = x_1 +0.5(x_1^2-0.21)x_2 &
 \end{array}\right.
\end{aligned} \quad 
\begin{aligned}
  \centering
         q_2:\left\{ \begin{array}{lr}
      \dot{x}_1  = -2 x_2, &\\
   \dot{x}_2 = 0.8x_1 +5(x_1^2-0.21)x_2 &
 \end{array}\right.
\end{aligned}
\end{equation*}

\begin{figure}[htbp]
\centering
\begin{minipage}[t]{0.48\textwidth}
\centering
  \begin{tikzpicture}
  \centering
\tikzstyle{mode} = [circle, minimum size=0.8cm,inner sep=0pt, text centered, draw=black, fill=white!70]
\node (mode1) [mode, align=center] { $q_1$};
\node (mode2) [mode, right of=mode1, xshift=3.2cm] {$q_2$};
\draw [->,thick,bend left]  (mode1) to   node[anchor=south, yshift=0.1cm] {$e_1=(q_1, q_2)$} (mode2);
\draw [->,thick]  (mode2) to   node[anchor=south, yshift=-0.1cm] {$e_2=(q_2, q_1)$} (mode1);
\end{tikzpicture}
\caption{ The HA for van  der Pol oscillator}
\label{fig:ex1}
\end{minipage}
\begin{minipage}[t]{0.48\textwidth}
\centering
  \begin{tikzpicture}
  \centering
\tikzstyle{mode} = [circle, minimum size=1cm,inner sep=0pt, text centered, draw=black, fill=white!70]
\node (mode1) [mode, align=center] { cruise};
\node (mode2) [mode, right of=mode1, xshift=3.2cm] { avoid};
\draw [->,thick]  (mode1) to   node[anchor=south, yshift=0.1cm] {$e=(q_1, q_2)$} (mode2);
\node [above of=mode1] {$q_1$};
\node [above of=mode2] {$q_2$};
\end{tikzpicture}
\caption{ The HA for aircraft conflict resolution}
\label{fig:ex4}
\end{minipage}
\end{figure}
In Fig.\ref{fig:ex1}, $\mathtt{Dom}_{q_1}=\mathtt{Dom}_{q_2}=\mathtt{Init}_{q_1}
 =\mathtt{Init}_{q_2}=\{ (x_1,x_2)  \mid x_1^2 +x_2^2 <1 \}$, and  
$\mathcal G_{e_1}  =  \{(x_1,x_2)  \mid 10 x_1^2+10 (x_2-0.3)^2<1 \}$ and 
$\mathcal G_{e_2} = \{(x_1,x_2)  \mid 15 x_1^2+10 x_2^2-1<0 \}$.
The system goal is  to guarantee 
the safety $\mathcal{S}_{q_1} = \{ (x_1,x_2)  \mid x_1^2 +x_2^2 <1 \wedge 10 x_1^2+10 (x_2-0.3)^2>1  \}$ and $\mathcal{S}_{q_2} = \{ (x_1,x_2)  \mid x_1^2 +x_2^2 <1 \wedge 15 x_1^2+10 x_2^2-1>0 \wedge 36 x_1^2+36 (x_2-0.7)^2-1>0 \}$. 

\begin{figure}[htp]
\centering
\subfigure[]{
\includegraphics[width=4cm]{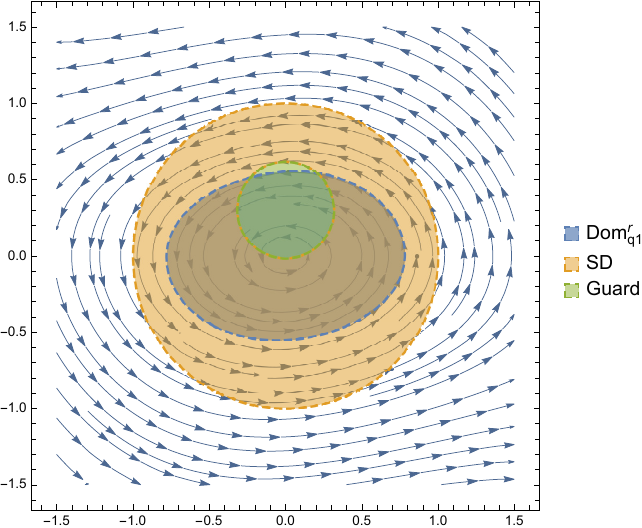}
}
\hspace{-2mm}
\subfigure[]{
\includegraphics[width=4cm]{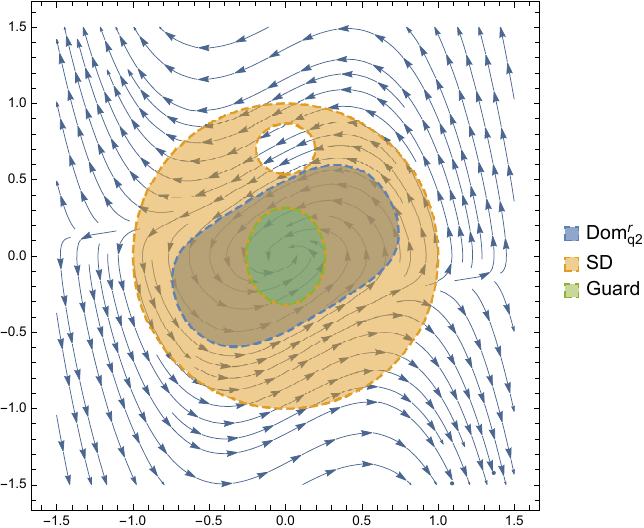}
\hspace{-2mm}
}
\caption{The blue regions in (a) and (b) represent the refined domains $\mathtt{Dom}^r_{q_1}$ and $\mathtt{Dom}^r_{q_2}$ respectively, and the green regions represent $\mathcal G_{e_1}$ and $\mathcal G_{e_2}$ respectively.  The synthesized reset controller is given by  $\mathtt{Init}^r_{q_1}=\mathtt{Dom}^r_{q_1}$ , $\mathtt{Init}^r_{q_2}=\mathtt{Dom}^r_{q_2}$ ,  $\mathcal{R}^r(e_1,x)\subset  \mathtt{Dom}^r_{q_2}$ and $\mathcal{R}^r(e_2,x)\subset \mathtt{Dom}^r_{q_1}$. 
}
\label{fig:resultEx2}
\end{figure}

A reset controller synthesised satisfying the system goal 
is illustrated in Fig. \ref{fig:resultEx2}. 

\subsection{An aircraft conflict resolution}\label{sec:aircraft}
The example of aircraft conflict resolution is taken from \cite{tomlin2000}, which 
models the kinematic motions of two aircrafts,  as showed in Fig.\ref{fig:ex4}. 
It has  two modes $q_1$, the cruise mode, and $q_2$, the avoid mode. 
$(x_1, x_2)$ and  $\Phi_r$ stand for 
the relative position and orientation of the two aircrafts, respectively.  
$\Phi_r$ is assumed to be constant.
$v_i$ for $i=1,2$ stand for the velocities  of the two  aircrafts, with  $-v_1+v_2\cos{\Phi_r}=0.1$ and $v_2\sin{\Phi_r}=-0.2$.

\noindent
The vector fields in  the two modes are 
\begin{equation*}
\begin{aligned}
         q_1:\left\{ \begin{array}{lr}
      \dot{x}_1  = -v_1 + v_2\cos{\Phi_r}, &\\
      \dot{x}_2 = v_2\sin{\Phi_r} &
 \end{array}\right.
\end{aligned} \quad 
\begin{aligned}
         q_2:\left\{ \begin{array}{lr}
      \dot{x}_1  =  -v_1 + v_2\cos{\Phi_r}+x_2. &\\
      \dot{x}_2 =v_2\sin{\Phi_r}-x_1 &
 \end{array}\right.
\end{aligned}
\end{equation*}
$\mathtt{Init}_{q_1}=\mathtt{Dom}_{q_1}=\{ (x_1,x_2)  \mid x_1^2 +x_2^2 <4 \}$, $\mathtt{Init}_{q_2}=\mathtt{Dom}_{q_2}=\{ (x_1,x_2)  \mid x_1^2 +x_2^2 <1 \}$, and  $\mathcal{G}_e=\{(x_1,x_2) \mid x_1^2+x_2^2<1 \}$.
\begin{figure}[htp]
 	\centering
 	\subfigure[]{
 		\includegraphics[width=4cm]{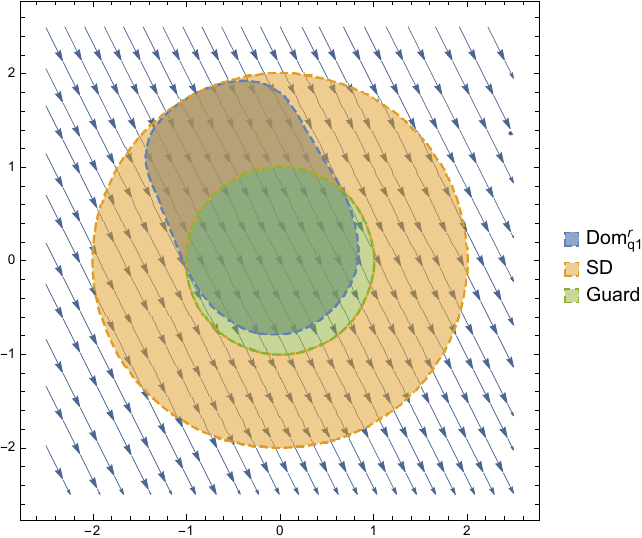}
 	}
 	\hspace{-2mm}
 	\subfigure[]{
 		\includegraphics[width=4cm]{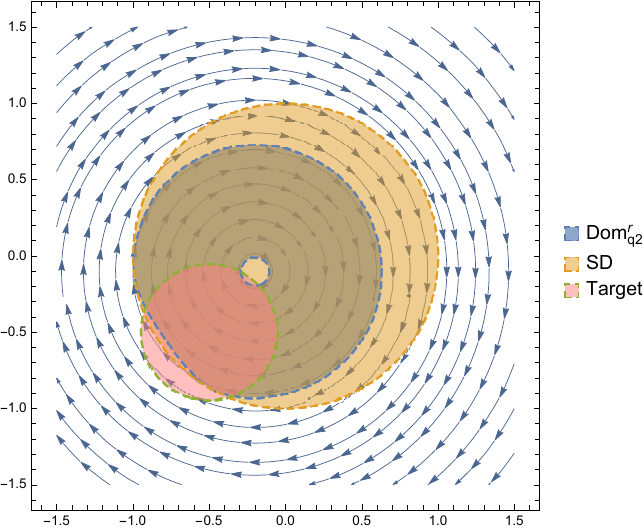}
 	}
 	\caption{The pink region in mode $q_2$  represents the target set.  
 	The blue regions of (a) and (b) represent the refined  domains $\mathtt{Dom}^r_{q_1}$, $\mathtt{Dom}^r_{q_2}$, respectively. The refined
 	initial sets are $\mathtt{Dom}^r_{q_1}$ and $\mathtt{Dom}^r_{q_2}$, 
 	and the reset map $\mathcal{R}^r(e,x)\subset\mathtt{Dom}^r_{q_2}$. }
 	\label{fig:result_aircraft}
 \end{figure}
The system goal is to reach the target set $\mathtt{Tr}=\{(q_2,x_1,x_2) \mid 5(x_1 + 0.5)^2 + 5(x_2 + 0.5)^2 < 1 \}$, 
meanwhile keep the safety  
$\mathcal{S}_{q_1}=\{(x_1,x_2) \mid x_1^2+x_2^2 < 4 \wedge x_1^2+x_2^2 > 1\}$ and  
$\mathcal{S}_{q_2}=\{ (x_1,x_2)  \mid x_1^2 +x_2^2 <1 \}$. 

A synthesized reset controller that satisfies the system goal  is 
presented pictorially in Fig.~\ref{fig:result_aircraft}. 

 \subsection{Clohessy-Wiltshire-Hill (CWH) equations}
 CWH equations describes a simplified model of the relative orbital motion of a chase spacecraft w.r.t. a target one \cite{7798767} as follows:
 \begin{align*}
          \dot{\bm{x}}
    =
     \left[ \begin{array}{cccc}
         0 & 1 & 0 & 0 \\
         3\omega^2 & 0 & 0 & 2\omega \\
         0 & 0 & 0 & 1 \\
         0 & -2\omega & 0 & 0
     \end{array}\right] \bm{x}
    +
     \left[ \begin{array}{cc}
          0 & 0 \\
          \frac{1}{m_c}& 0 \\
          0 & 0\\
          0 & \frac{1}{m_c}
     \end{array}\right]
    \bm{u}
 \end{align*}
 where $\bm{x}=[x_1,x_2,x_3,x_4]\in \mathbb{R}^4$ are state variables, $\bm{u}=[u_1,u_2]$ 
are inputs, $\omega$ is the mean motion of the reference orbit, and $m_c$ 
is the mass of the chase spacecraft. 
 \begin{wrapfigure}{r}{0.7\textwidth}
\centering \vspace*{-6mm} 
\scalebox{0.85}{
  \begin{tikzpicture}
  \centering
\tikzstyle{mode} = [circle, minimum size=2.2cm,inner sep=0pt, text centered, draw=black, fill=white!60]
\node (mode1) [mode, align=center] {};
\node (mode2) [mode, below of=mode1, xshift=-2.6cm, yshift=-3cm]{ };
\node (mode3) [mode, below of=mode1, xshift=2.6cm, yshift=-3cm] {};
\draw [->,thick]  (mode1) to    node[anchor=south, xshift=-0.9cm,  yshift=0cm] {$e_1=(q_1, q_2)$} (mode2);
\draw [->,thick] (mode2) to  node[anchor=south,  yshift=0.0cm] {$e_2=(q_2, q_3)$} (mode3);
\draw [->,thick] (mode3) to node[anchor=south, xshift=0.9cm, yshift=0cm]{$e_3=(q_3, q_1)$} (mode1);
\node [above of=mode1, yshift=-1.0cm] {$ \begin{array}{c} 
q_1\\
m_c=150;\\
\bm{u}=[1,1];\\
\omega=1
\end{array}$};
\node [above of=mode2, yshift=-1.0cm] {$ \begin{array}{c} 
q_2\\
m_c=150;\\
\bm{u}=[3,3];\\
\omega=1.5
\end{array}$};
\node [above of=mode3, yshift=-1.0cm] {$ \begin{array}{c} 
q_3\\
m_c=150;\\
\bm{u}=[-1,-1];\\
\omega=1
\end{array}$};
\node [above of=mode3, xshift=-1.5cm,yshift=-2.8cm] {$ \begin{array}{rcl} 
\mathcal{G}_{e_1}&=&\{x\mid (x_1+1)^2+(x_2+1)^2+x_3^2+x_4^2<1\};\\
\mathcal{G}_{e_2}&=&\{x\mid (x_1-1)^2+(x_2-1)^2+x_3^2+x_4^2<1\};\\
\mathcal{G}_{e_3}&=&\{x\mid (x_1-1)^2+(x_2+1)^2+(x_3+1)^2+x_4^2<1\}
\end{array}$};
\end{tikzpicture}}
\caption{ The HA for the CWH equations with three modes}
\label{fig:CWH}
\end{wrapfigure}
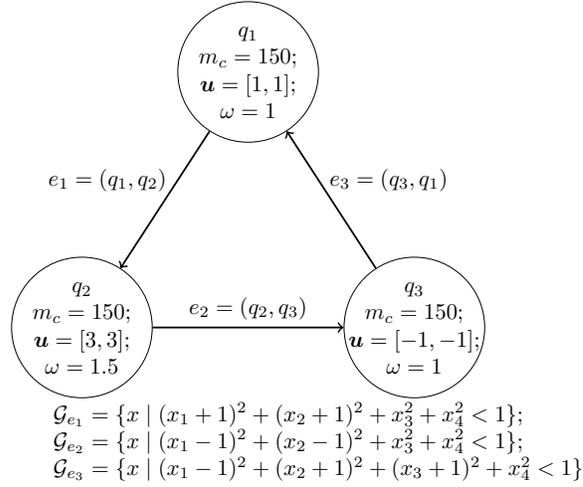
The whole system consists of the three modes with different values of $u$, 
$\omega$, and $m_c$, which forms  a circle as shown in Fig~\ref{fig:CWH}.
In which, the domain, initial set, and safe set of each mode are all unit hyperspheres.  Suppose $\mathtt{TR}=\{(q_3,x_1,x_2,x_3,x_4) \mid (x_1+1)^2+(x_2-1)^2+x_3^2+x_4^2<1\}$, the system goal is
to guarantee the reachable of the target meanwhile keeping the 
safety.  

\begin{figure}[htp]
 	\centering
 	\subfigure[]{
 		\includegraphics[width=3.6cm]{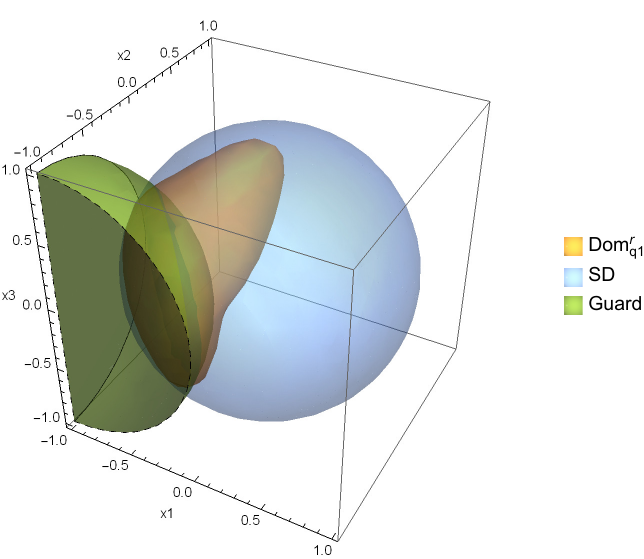}
 	}
 	\hspace{-3mm}
 	\subfigure[]{
 		\includegraphics[width=3.6cm]{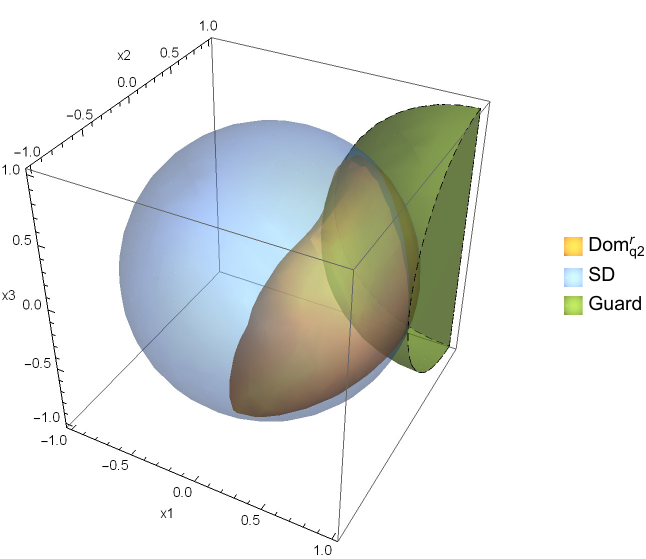}
 	}
 	\subfigure[]{
 		\includegraphics[width=3.6cm]{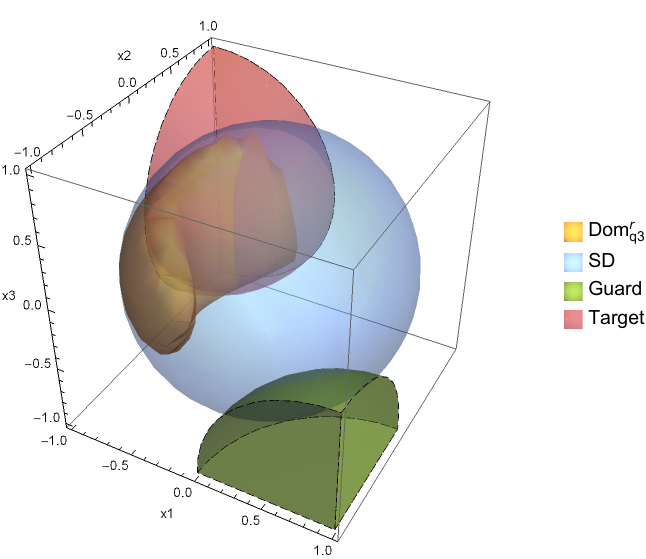}
 	}
 	\caption{An illustration of three modes of the CWH equations with $x_4=0$. 
 	The green regions stands for $\mathcal{G}_{e_i}$
 	 for $i=1,2,3$, and the pink region in (c) represents  the target set $\mathtt{TR}$. The orange regions represent the refined domain  $\mathtt{Dom}_{q_i}^r$ computed by our approach, the refined 
 	 initial set $\mathtt{Init}_{q_i}^r$ is the same as $\mathtt{Dom}_{q_i}^r$ 
 	 for $i=1,2,3$, and the synthesized reset maps are  $\mathcal{R}^r(e_1,x)\subset\mathtt{Dom}^r_{q_2}$, $\mathcal{R}^r(e_2,x)\subset\mathtt{Dom}^r_{q_3}$ and $\mathcal{R}^r(e_3,x)\subset\mathtt{Dom}^r_{q_1}$. }
 	\label{fig:cw}
 \end{figure}

A synthesized reset controller satisfying the system goal   
is presented pictorially in Fig.~\ref{fig:cw}. 
To avoid the unreachability caused by 
 the loops consisting of $q_1,q_2$ and $q_3$, 
 the synthesized reset controller blocks them at the edge $e=(q_3,q_1)$.
 
\newcolumntype{F}{>{\quad}c<{\quad}}
\begin{table}[htp]
    \centering
    \begin{tabular}{c|F F F}
        \hline Cases & \multicolumn{3}{|c }{Time Consuming of each mode}  \\
        \hline Van der Pol oscillator & 2.602795 & 1.537758 & \\
        \hline Aircraft Conflict Resolution & 1.218457 & 1.397034 & \\
        \hline Clohessy-Wiltshire-Hill Equations & 72.033725 & 79.730631 & 95.515169\\
        \hline
    \end{tabular}
    \caption{This table shows the time consuming of our three case studies. The running time of \ref{alg:reachavoid} for each modes is counted in second and presented with individual columns }
    \label{tab:time_consuming}
\end{table}

 \section{Conclusion}
\label{sec:conclusion}
In this paper, motivated by practice, we investigated the reset controller synthesis problem for polynomial HSs subject to safety and liveness. We first gave a sound and complete method to synthesizing reset controllers for polynomial HSs only with safety property, which was by reduction to reach-avoid and differential invariant problems that can be solved by semi-define programming. Then we took liveness property into account. This is much involved in order to guarantee the reachability of the target set but is essentially solved by reduction to reach-avoid problems in each mode and searching all simple loops in the directed graph consisting of the modes and the edges of the considered HS to cope with the unreachability caused by infinite loops among modes. We implemented a prototypical tool and applied it to case studies, including one from the real-world spacecraft scenario, i.e., CWH equations, which indicate the effectiveness and efficiency of our approach.

In the future work, we would like to extend our approach to HSs with delays. In addition, it deserves to investigate correct-by-construction for HSs by taking feedback controller synthesis, switching logic controller synthesis, and reset controller synthesis into account uniformly.
\bibliographystyle{abbrv}
\bibliography{reference}

\newpage

\appendix
\section{Proof of Lemma \ref{lem:subcon} }
Suppose  $\bm{x} \in \mathtt{RA}^*(\mathcal X_0\xrightarrow[f]{S} \mathtt{TR})$,  then there is $T\geq 0$ such that (1) ${\phi}(\bm{x},t)\in  S$ for all $t\in[0,T)$; 
(2) for any $\epsilon >0$ there is some $t \in [T, T+\epsilon)$ with  ${\phi}(\bm{x},t )\in  \mathtt{TR}$. 
As $S\cap \mathtt{TR}=\emptyset$, we have $\phi(\bm{x},T)\in\partial S$, then
 $\phi(\bm{x},T) \in \mathtt{trans}^*_{\bm{f}\uparrow S}$. Thus, 
 ${\phi}(\bm{x},T)\in \bar{S} \cap \overline{\mathtt{TR}}$. 
 Therefore, for an $\epsilon>0$, there must be $t < T$ such that ${\phi}(\bm{x},t) \in \mathcal B({\phi}(\bm{x},T), \epsilon/2) =\{\bm{y} \mid ||\bm{y}-{\phi}(\bm{x},T)|| < \epsilon/2\}\subseteq \mathcal N_{\epsilon}(\mathtt{TR})$. Hence, $\bm{x} \in \mathtt{RA}(S, f, \mathtt{TR}_{\epsilon}(S))$. Therefore, $ \mathtt{RA}^*(\mathcal X_0\xrightarrow[f]{S} \mathtt{TR}) \subseteq  \bigcap\limits_{\epsilon >0} \mathtt{RA}(S,f, \mathtt{TR}_{\epsilon}(S))$.
 
 \section{Proof of Theorem \ref{thm:RAPT} }
 For ``$\subseteq$'', suppose 
$\bm{x}\in \mathtt{RA}^*( S \xrightarrow[f]{S}  \mathtt{trans}^*_{{f}\uparrow S})$. Then,  there is $T_{\bm{x}}\geq 0$ such that ${\phi}(\bm{x},T_{\bm{x}})\in \mathtt{trans}^*_{ f\uparrow S}$ and ${\phi}(\bm{x},t)\in S, \forall t\in [0,T_{\bm{x}})$. 
Clearly, $\bm{x} \notin \infty\!\!-\!\!\textit{Safe}( \bm{f}, S,  S)$. 
So, we only need to show $\bm{x}\in \bigcap\limits_{\epsilon >0} \mathtt{RA}(S, f,  (\mathtt{trans}^*_{\bm{f}\uparrow S})_\epsilon(S))$, which 
can be done by considering the following two cases: When  
 ${\phi}(\bm{x},T_{\bm{x}})\in S$, it follows 
 ${\phi}(\bm{x},T_{\bm{x}})\in \bar{\partial}S \cap \mathtt{trans}^*_{ f\uparrow S }$.  Thus, for any $\epsilon >0$,  ${\phi}(\bm{x},T_{\bm{x}})\in  \mathtt{RA}(S,  f,  (\mathtt{trans}^*_{\bm{f}\uparrow S})_\epsilon(S))$.
When ${\phi}(\bm{x},T_{\bm{x}})\in \partial^\circ  S$, 
 it follows that 
  $\{{\phi}(\bm{x},t)\mid 0 \leq t <T_{\bm{x}} \} \cap  (\mathtt{trans}^*_{\bm{f}\uparrow S})_\epsilon(S) \neq \emptyset$  for any $\epsilon >0$. So,  ${\phi}(\bm{x},T_{\bm{x}})\in  \mathtt{RA}(S,  f,  (\mathtt{trans}^*_{\bm{f}\uparrow S})_\epsilon(S))$. 
  \oomit{Thus, 
\[\mathtt{RM} \subseteq  \textit{LRA}( S \xrightarrow[ f]{ S} \mathtt{trans}^*_{ f\uparrow S}) -ES( f, S,  S).\] }

For ``$\supseteq$'', 
suppose \[\bm{x}\in \bigcap\limits_{\epsilon >0} \mathtt{RA}(S, f,  (\mathtt{trans}^*_{\bm{f}\uparrow S})_\epsilon(S)) \setminus  \infty\!\!-\!\!\textit{Safe}(\bm{f}, S, S).\] Then, there is a $T_{u}\in [0,\infty)$, such that $\phi(\bm{x},T_{u})\notin S$. Hence, the greatest lower bound of $\{t\mid \phi(\bm{x},t)\notin S\}$ exists, set $T_x = \inf{\{t \mid\phi(\bm{x},t)\notin S\} }$. Then,  $\forall t\in[0,T_x)$, $\phi(\bm{x},t)\in S$. Besides, $\forall \epsilon >0$, $\exists t\in [T_x,T_x+\epsilon)$, $\phi(\bm{x},t)\notin S$, which is to say, $\phi(\bm{x},T_x)\in \mathtt{trans}^*_{\bm{f}\uparrow S}$, so $\bm{x}\in \mathtt{RA}^*( S \xrightarrow[f]{S}  \mathtt{trans}^*_{{f}\uparrow S})$

\section{Proof of Theorem \ref{thm:DI} }

For any $\bm{x}\in S$, if there is some $T_{\bm{x}}\geq 0$ such that ${\phi}(\bm{x},t)\in  S$ for all $t\in [0, T_{\bm{x}})$, but ${\phi}(\bm{x},T_{\bm{x}})\notin  S$, then ${\phi}(\bm{x},T_{\bm{x}})\in \mathtt{trans}^*_{ f\uparrow  S}$. Hence  $\bm{x}\in\mathtt{RA}^*( S \xrightarrow[f]{S}  \mathtt{trans}^*_{{f}\uparrow S})$. Additionally, if there is some $T_{\bm{x}}\geq 0$ such that ${\phi}(\bm{x},t)\in  S$ for all $t\in [0, T_{\bm{x}}]$, but for any $\epsilon>0$ there is a $t_\epsilon \in (T_{\bm{x}}, T_{\bm{x}}+\epsilon)$ such that ${\phi}(\bm{x},t_\epsilon)\notin  S$, then ${\phi}(\bm{x},T_{\bm{x}})\in \mathtt{trans}^*_{ f\uparrow  S}$.  Whence  $\bm{x}\in \mathtt{RA}^*( S \xrightarrow[f]{S}  \mathtt{trans}^*_{{f}\uparrow S})$. Therefore, it follows that 
if $\bm{x}\in C$ then  for arbitrary $T$, ${\phi}(\bm{x},t)\in  S$ for all $t\in [0,T]$ implies ${\phi}(\bm{x},t)\in  C$ for all $t\in [0,T]$.  As a result, $ C $ is  a DI of  $ f$ w.r.t. $S$.

\section{Proof of Theorem \ref{thm:ssf} }
For soundness,
	any $\bm{x}_q \in \mathtt{Init}^r_q$ , $\bm{x}_q$ is either in the invariant set of $\mathtt{SD}_q$ or there is some edge $e=(q, p)\in\mathcal E$ such that  $\bm{x}_q$ safely arrives $\mathtt{Dom}_q^c \cap \mathcal{G}_e$.
	
	In the first case, the whole trajectory $\bm{\phi}(q,\bm{x},\cdot)$ in $\mathtt{SD}_q$ by Theorem \ref{thm:DI} and so in $\mathcal{S}_q $. While in the second case, there is some $T\geq 0$ such that $\bm{\phi}(q,\bm{x},\cdot)$ safely reaches  $\bm{\phi}(q,\bm{x},T)$ and either $\bm{\phi}(q,\bm{x},T)\in \mathtt{Dom}_q^c \cap \mathcal{G}_e $ or for any $\epsilon >0$ there is some $t_\epsilon\in (T, T+\epsilon)$ such that $\bm{\phi}(q,\bm{x},t_\epsilon)\in \mathtt{Dom}_q^c \cap \mathcal{G}_e$. In any case, a discrete jump to $p$ will occur no later than $T$ and ensure safety before the jump.
	
	By the construction, the image of any reset map $\mathcal R(e, \bm{x})$ with $e=(w, q)$ must be in  $\mathtt{Dom}^r_{q}$. By a similar argument to the last paragraph,  the trajectory $\bm{\phi}(q,\bm{x},\cdot)$  for any $\bm{x}\in \mathtt{Dom}^r_{q}$ must either stay in $\mathtt{SD}_{q}$ forever or $\mathtt{SD}_{q}$-safely reaching some state $\bm{\phi}({q},\bm{x}, T_{\bm{x}})$ at which a discrete jump occurs. In any case, the system does not block at $q$.
	Therefore,  each trace $(\bm{\sigma}, \bm{\phi}, \bm{\tau})$ starting with any $(q, \bm{x})\in \mathtt{Init}^r$ is infinite and always stays in  $\mathcal S$. So,  $\mathcal H^r $ solves \textbf{Problem I}.

    For completeness, if \textbf{Problem I} can be solved by some reset controller, then the trace of system is either in the invariant of some mode or infinitely jumping between different mode, in neither of which $\mathtt{Dom}^r=\bigcup_{q\in\mathcal{Q}}\mathtt{Dom}^r_q$ is empty, so there will be some mode $q$ satisfying that $\mathtt{Init}^r_q=\mathtt{Init}_q\cap\mathtt{Dom}^r_q\neq\emptyset$.

\section{Proof of Theorem \ref{thm:lsf} }

For the soundness, assume there is a  trace $(\bm{\sigma}, \bm{\phi}, \bm{\tau})$ satisfying the configuration of  $\mathcal{H}^r$ outputted by Algorithm \ref{alg:LivResetSyn} but never $\mathcal S$-safely reaching into any $\mathtt{TR}_{q}$ in finite time. As line \ref{alg3:line1}-\ref{alg3:line9} of Algorithm \ref{alg:LivResetSyn} are the same as Algorithm \ref{alg:SafeResetSyn} (the only difference is that for any $x\in\{y\mid (q,y)\in\mathtt{Init}_q^r\vee y\in\mathcal{R}^r(e,y')\,,\,q\in\mathcal{Q}\,,\,e\in\mathcal{E}\}$, the trajectories starting from $x$ will only enter some state confirm the next jump rather, not stay in the invariant of current mode). By Theorem \ref{thm:ssf}, all the traces generated by Algorithm \ref{alg:LivResetSyn} maintain safety before termination, which includes $(\bm{\sigma}, \bm{\phi}, \bm{\tau})$.  Then it contains some loops of modes that maintain the safety property infinitely without reaching the target set.  

As the depth-first searching method of Algorithm \ref{alg:DomRef} will explore all the trace of $\mathcal{H}$, $(\bm{\sigma}, \bm{\phi}, \bm{\tau})$ will be explored. Without loss of generation, assuming $\bm{\sigma}=(q_0,q_1,\cdots,$ $(q_i,\cdots,q_n)^{\omega})$, where $(\cdot)^{\omega}$ means an infinite loop of the sequence in brackets. Then we can know that $\mathtt{Post}(q)$ of any mode in $\bm{\sigma}$ can not be $\emptyset$, so the program will always enter the else part of Algorithm \ref{alg:DomRef} (line \ref{alg2:line17}) and stop exploring new mode when the second $q_i$ is added to $\textit{Path}$. Then if $\mathtt{ST}_{q_i}\neq\emptyset$, the reset map from $q_n$ to $q_i$ will be refined so that the trace will enter a target set at the second time it enter mode $q_i$, otherwise the jump from $q_n$ to $q_i$ will be blocked by refine $\mathtt{Dom}_{q_n}$, either cases will break the loop.

For the completeness, assume a hybrid automata $\mathcal{H}$ can be refined by a reset controller $\mathcal{H}^r$ to satisfy some safety property together with liveness. Then any trace in $\mathcal{H}^r$ will eventually enter a target set in a mode $q$ without any probability to enter guard conditions in $q$, which is to say, the reset map of the edge jump to $q$ is a subset of $\mathtt{ST}_q$. Meanwhile, the possible jumps in any mode of $\mathcal{H}^r$ are confirmed to enter such a mode $q$ within finite time. The Depth-First Searching method confirm all the traces of $\mathcal{H}$ are searched and all the jump $e=(p,q)$ leading to some infinite safe loop or some block mode are broke down by refine the reset map $\mathcal{R}^r((\mathtt{Pre}(p),p),x)$ as a subset of $\mathtt{Dom}_p\setminus\mathtt{RA}^{*} (\mathtt{SD}_p\xrightarrow[\bm{f}_p]{ \mathtt{SD}_p}\mathcal{G}_{e})$. 

\section{Proof of Theorem \ref{thm:IAP} }

Let $\bm{x}\in \mathtt{IAP}$. Obviously, $\bm{x}\in \mathcal{S}$ by the definition of $\mathtt{IAP}$. If $\bm{x}\in  \mathtt{TR}$ then the conclusion holds obviously. In the following  we will prove that if $\bm{x}\in \mathcal{S}  \setminus \mathtt{TR}$, then there must exist some $t\in \mathbb{R}_{\geq 0}$ such that 
\begin{equation}\label{eq:toproof}
	\bm{\phi}(\bm{x},t) \in \mathtt{TR} \wedge \bigwedge_{\tau \in [0,t)}\bm{\phi}(\bm{x},\tau) \in \mathcal{S}
\end{equation}

Assume \eqref{eq:toproof} does not hold, then there are two possibilities:
\begin{enumerate}
		\item  There exists $t\in (0,\infty)$ such that 
		\begin{equation}\label{eq:contr1}
		\bm{\phi}(\bm{x},t) \in \partial \mathcal{S} \wedge \bigwedge_{\tau\in [0,t)}\bm{\phi}( \bm{x},\tau) \in \mathcal{S} \setminus \mathtt{TR}.
		\end{equation}
	
		\item For any $\tau \in \mathbb{R}_{\geq 0}$, 
		\begin{equation}
		\label{eq:infintelysafe}
		\bm{\phi}(\bm{x},\tau) \in \mathcal{S} \setminus \mathtt{TR}.
		\end{equation}	
		
	\end{enumerate}
	
	Assume \eqref{eq:contr1} true. As a result, 	
	\begin{equation}
	\label{eq:deduceresult}
	\theta(\bm{\phi}(\bm{x},\tau))\leq \theta(\bm{x})<0, \forall \tau \in [0,t),
	\end{equation}
	implies that $\theta(\bm{\phi}(\bm{x},t))<0$, contradicting the fact that $\theta(\bm{x})\geq  0$ for $\bm{x}\in \partial \mathcal{S}$. Therefore, \eqref{eq:contr1} cannot hold.

Now assume that \eqref{eq:infintelysafe} holds. 
	From \eqref{eq:racon1}, we have that
	\begin{equation}
	\label{mono1}
	\theta(\bm{\phi}(\bm{x},\tau))\leq \theta(\bm{x})<0, \forall \tau \in \mathbb{R}_{\geq 0},
	\end{equation}
	From \eqref{eq:racon2}, we have that for $\tau \in \mathbb{R}_{\geq 0}$,
	\begin{equation}
	\label{ineq1}
	\begin{split}
	\theta(\bm{\phi}(\bm{x},\tau))\geq &  \max\limits_{k=1,2,\cdots,K}\gamma_k(\bm{\phi}(\bm{x},\tau) ) + \left\langle \frac{\partial \psi(\bm{x})}{\partial \bm{x}},  \bm{f}(\bm{x}) \right\rangle \bigg|_{\bm{x}=\bm{\phi}(\bm{x},\tau)}.
	\end{split}
	\end{equation}
	Thus, 
	\begin{equation}
	\label{ineq}
	\begin{split}
	0> \max\limits_{k=1,2,\cdots,K}\gamma_k(\bm{\phi}(\bm{x},\tau) ) + \left\langle  \frac{\partial \psi(\bm{x})}{\partial \bm{x}}, \bm{f}(\bm{x})\right\rangle\bigg|_{\bm{x}=\bm{\phi}(\bm{x},\tau)}.
	\end{split}
	\end{equation}
	We have $ \left\langle\frac{\partial \psi(\bm{x})}{\partial \bm{x}}, \bm{f}(\bm{x})\right\rangle\bigg|_{\bm{x}=\bm{\phi}(\bm{x},\tau)}<-\max\limits_{k=1,2,\cdots,K}\gamma_k(\bm{\phi}(\bm{x},\tau) )<0$ for $\tau \in \mathbb{R}_{\geq 0}$ since $	\bm{\phi}(\bm{x},\tau) \notin  \mathtt{TR}$ and consequently $\psi(\bm{\phi}(\bm{x},\tau))< \psi(\bm{x}), \forall \tau \in \mathbb{R}_{\geq 0}.$ Also, since $\psi(\bm{\phi}(\bm{x},\tau))$ is bounded for $\tau \in \mathbb{R}_{\geq 0}$, we have \[\lim_{\tau_1\rightarrow \infty, \tau_2\rightarrow \infty}(\psi(\bm{\phi}(\bm{x},\tau_1))-\psi(\bm{\phi}(\bm{x},\tau_2))=0.\] Further,  we have 
	\begin{equation}
	\label{zero}
	\lim_{\tau \rightarrow \infty} \left\langle \frac{\partial \psi(\bm{x})}{\partial \bm{x}},\bm{f}(\bm{x})\right\rangle\bigg|_{\bm{x}=\bm{\phi}(\bm{x},\tau)}=0.
	\end{equation}
	However, from   \eqref{mono1}-\eqref{ineq}, we have that for $\tau \in \mathbb{R}_{\geq 0}$,
	\begin{equation*}
	\begin{split}
	&\left\langle\frac{\partial \psi(\bm{x})}{\partial \bm{x}},\bm{f}(\bm{x})\right\rangle\bigg|_{\bm{x}=\bm{\phi}(\bm{x},\tau)}\leq \theta(\bm{x}) - \max\limits_{k=1,2,\cdots,K}\gamma_k(\bm{\phi}(\bm{x},\tau) ) \leq \theta(\bm{x})
	\end{split}
	\end{equation*}
	which contradicts \eqref{zero} since $\theta(\bm{x})<0$ by \eqref{mono1}. 
	
	Therefore, $\mathtt{IAP} = \{\bm{x} \in  \mathcal{S} \mid \theta(\bm{x})<0\}$ is an inner-approximation of the reach-avoid set $\mathtt{RA}^*(\mathcal S \xrightarrow[f]{\mathcal S} \mathtt{TR})$.
	
	As a result, for $\bm{x}\in\mathtt{IAP}$, the trajectory $\bm{\phi}(\bm{x},\cdot)$ will $\mathcal{S}$-safely enter $ \mathtt{TR}$. As a consequence, the trajectory $\bm{\phi}(\bm{x},\cdot)$ starting with any $\bm{x}\in\mathtt{IAP}^*=\mathcal X_0 \cap \mathtt{IAP}$ will $\mathcal{S}$-safely enter $ \mathtt{TR}$. Therefore, $\mathtt{IAP}^*=\mathcal X_0 \cap \mathtt{IAP}$ is an inner-approximation of  $\mathtt{RA}^*(\mathcal X_0\xrightarrow[f]{\mathcal S} \mathtt{TR})$.

\section{Proof of Theorem \ref{inner0} }

	Obviously, the fact that $\theta(\alpha^*, \bm{x})$ satisfies constraints in \eqref{sos} implies that $\theta(\alpha^*, \bm{x})$ satisfies \eqref{eq:racon1}-\eqref{eq:racon3} according to $\mathcal{S}$-procedure in \cite{boyd1994}. Therefore, the value returned by  Algorithm \ref{alg:reachavoid} is an inner-approximation  as Theorem \ref{thm:IAP}.

\end{document}